\documentclass[aps,
 reprint,
 superscriptaddress,
 showpacs,
 amsmath,amssymb,
 aps,
]{revtex4-1}

\usepackage[normalem]{ulem}
\usepackage{graphicx}
\usepackage[normalem]{ulem}
\usepackage{dcolumn}
\usepackage{bm}
\usepackage{hyperref}
\usepackage{bbold}

\usepackage{epstopdf}
\usepackage{todonotes}
\usepackage{siunitx}
\usepackage{units}

\newcommand{\idmat}{\ensuremath{\mathbb{1}}}

\providecommand{\Eprint}[2]{}

\begin{document}

\title{Interplay between charge, magnetic and superconducting order in a Kondo lattice with an attractive Hubbard interaction}
\author{Benedikt Lechtenberg}
\author{Robert Peters}
\author{Norio Kawakami}
\affiliation{Department of Physics, Kyoto University, Kyoto 606-8502, Japan}

\date{\today}

\begin{abstract}
We investigate the competition between superconductivity, charge-ordering, magnetic-ordering, and the Kondo effect in a heavy fermion $s$-wave superconductor
described by a Kondo lattice model with an attractive on-site Hubbard interaction.
The model is solved using the real-space dynamical mean field theory. 
For this purpose, we develop a numerical renormalization group (NRG) framework in Nambu space,
which is used to solve the superconducting impurity problem. 
This extended NRG scheme also allows for SU(2) spin symmetry broken solutions, 
enabling us to examine the competition or cooperation between $s$-wave superconductivity and incommensurate spin-density waves (SDWs).
At half filling, we find an intriguing phase where the magnetic ordering of the $f$-electrons lifts the degeneracy between the charge density wave (CDW) state and the superconducting state,
leading to a strong suppression of superconductivity.
In addition, the system may also become a half metal in this parameter regime.
Away from half filling, the CDWs vanish and are replaced by superconductivity combined with incommensurate SDWs up to moderate Kondo couplings to the $f$-electrons.
We find that both CDWs as well as superconductivity enhance magnetic ordering due to the suppression of Kondo screening.
\end{abstract}


\maketitle

\section{Introduction}
\label{sec:Introduction}
Strongly correlated electron systems attract enormous attention because of the multitude of remarkable phenomena they exhibit, 
such as the Kondo effect, magnetic or charge ordering, and unconventional superconductivity. 
The situation becomes particularly interesting when different effects are either competing or reinforcing each other. 
A class of compounds that exhibit all these effects are the heavy fermion materials 
\cite{lit:Hewson:KondoProblem93,lit:Grewe1984,lit:Stewart1984,lit:Stewart2001,lit:Stewart2006,lit:Lohnensysen2007,lit:Barzykin2005,lit:Si2001,lit:Si2006,lit:Coleman2005,lit:Gegenwart2008,lit:Knafo2009,lit:Si2010,lit:Paschen2014},
where strongly interacting $f$-electrons hybridize with conduction $spd$ bands.

Heavy fermion superconductors are usually considered to be a nodal unconventional superconductor 
where the nonlocal Cooper pairing is mediated by magnetic fluctuations \cite{lit:Ott1983, lit:Stewart1984_2,lit:Pfleiderer2009,lit:Nunes2003,lit:Steglich2016,lit:Masuda2015}.
However, very recently the pairing mechanism of the first heavy fermion superconductor $\mathrm{Ce}\mathrm{Cu}_2\mathrm{Si}_2$ is controversially discussed 
\cite{lit:Kittaka2014,lit:Yamashitae2017,lit:Kitagawa2017,lit:Takenaka2017,lit:Li2018,lit:Pang2018}.
While $\mathrm{Ce}\mathrm{Cu}_2\mathrm{Si}_2$ was generally believed to be a prototypical $d$-wave superconductor \cite{lit:Steglich1979}, 
recent low-temperature experiments have found no evidence of gap nodes at any point of the Fermi surface \cite{lit:Yamashitae2017}.
These results indicate that, contrary to the long-standing belief, $\mathrm{Ce}\mathrm{Cu}_2\mathrm{Si}_2$ is a heavy-fermion superconductor 
with a fully gapped $s$-wave superconducting (SC) state which may be caused by an on-site attractive pairing interaction.

Since it is generally believed that the coupling between conduction electrons and strongly interacting $f$-electrons, 
which causes the Kondo effect and magnetism, strongly suppresses superconductivity, 
heavy fermion superconductors with an attractive on-site pairing interaction 
have been barely studied theoretically \cite{lit:Bertussi2009,lit:Bodensiek2010,lit:Maska2010,lit:Bodensiek2013,lit:Karmakar2016,lit:Costa2018}. 
Furthermore, besides the possibility of fully gapped superconductivity in $\mathrm{Ce}\mathrm{Cu}_2\mathrm{Si}_2$, 
$s$-wave superconductivity might always be induced in heavy fermion systems via the proximity effect \cite{lit:Otop2003,lit:Chen1992,lit:Han1985}, 
making it possible to study the interplay between superconductivity,
magnetic ordering, charge ordering, and the Kondo effect.

One of the simplest models comprising all these effects is a Kondo lattice
\cite{lit:Doniach77,lit:Lacroix1979,lit:Fazekas1991,lit:Assaad1999,lit:Peters2007}
with an additional attractive Hubbard interaction $U<0$ \cite{lit:Bauer2009}:
\begin{align}
	H=& t\sum_{<i,j>,\sigma} \left( c^\dagger_{i,\sigma} c^{\phantom{\dagger}}_{j,\sigma} + \mathrm{H.c.} \right) - \mu \sum_{i,\sigma} n_{i,\sigma} \nonumber \\ 
	    &+ U\sum_i n_{i,\uparrow} n_{i,\downarrow} + J \sum_{i} \vec{S}_i \cdot \vec{s}_i, \label{eq:lattice_H}
\end{align}
where $\mu$ is the chemical potential, $t$ denotes the hopping parameter between nearest neighbors and $J>0$ is a Kondo coupling.
$c^\dagger_{i,\sigma}$ creates a conduction electron on site $i$ with spin $\sigma$ and $n_{i,\sigma} = c^\dagger_{i,\sigma} c^{\phantom{\dagger}}_{i,\sigma}$.
The last term in Eq. \eqref{eq:lattice_H} describes the spin-spin interaction 
between the conduction electron spins $\vec{s}_i=\sum_{\sigma,\sigma'}c^\dagger_{i,\sigma} \vec{\sigma}_{\sigma,\sigma'} c^{\phantom{\dagger}}_{i,\sigma'}$ 
and the localized $f$-electron spins $\vec{S}_i$,
with the Pauli matrices $\vec{\sigma}_{\sigma,\sigma'}$. 

This model has been investigated in one dimension by means of density matrix renormalization group (DMRG) for a filling of $n=1/3$ \cite{lit:Bertussi2009},
for different fillings in three dimensions using static mean-field theory \cite{lit:Costa2018} 
and for ferromagnetic couplings $J<0$ in two dimensions with the aid of variational minimization and Monte Carlo methods \cite{lit:Karmakar2016}.
For $U=0$, the model reduces to the ordinary Kondo lattice model, exhibiting a competition between spin-density waves (SDWs) and the Kondo effect,
while for $J=0$ one obtains the attractive Hubbard model with an on-site pairing term.
This on-site pairing may evoke superconductivity and, at half filling, also a charge density wave (CDW) state which is energetically degenerate with the SC state \cite{lit:Huscroft1997,lit:Bauer2009}.
Although CDWs play a crucial role at half filling, previous investigations of the model Eq. \eqref{eq:lattice_H} have ignored possible CDWs \cite{lit:Karmakar2016,lit:Costa2018}.
A finite $J$ and attractive $U$ allows us to examine the interplay between all these effects.
Such an attractive on-site term can arise in different ways.
In solid state systems, it can be mediated by bosons, e.g., phonons \cite{lit:Raczkowski2010} or excitons, while
in ultracold atom systems \cite{lit:Review_UltraColdGases2008} the effective interaction 
between optically trapped fermionic atoms can be tuned using Feshbach resonances \cite{lit:Tiesinga1993,lit:Stwalley1976,lit:Inouye1998,lit:Courteille1998}
so that it is well described by a local attractive potential.
In such systems, $s$-wave superfluidity has already been observed \cite{lit:Greiner2003,lit:Zwierlein2004,lit:Zwierlein2005,lit:Chin2006}.

In this paper, we investigate the interplay between 
magnetic ordering, charge ordering, the Kondo effect, and superconductivity for the Kondo lattice Hamiltonian 
with an attractive Hubbard interaction [Eq. \eqref{eq:lattice_H}] on a two-dimensional square lattice. 
To analyze this system, we employ the real-space dynamical mean field theory (RDMFT) which is a generalization of the dynamical mean-field theory (DMFT) \cite{lit:Metzner89,lit:DMFTreview}.
The DMFT has been proven to be very suitable to investigate the properties of strongly correlated lattice systems 
in cases where the momentum dependence of the self-energy can be neglected.
In the RDMFT, each lattice site of a finite cluster is mapped onto its own impurity model.
This allows us to study incommensurate CDWs or SDWs,
however, nonlocal interactions such as intersite SC pairing mechanisms cannot be described with the RDMFT.
Therefore, only $s$-wave superconductivity, mediated by a local pairing, is investigated in this paper.
The effective impurity models have to be solved self-consistently. 
For this purpose we develop a new self-consistent NRG \cite{lit:WilsonNRG,lit:BullaReview} scheme 
which allows us to combine superconductivity with spin symmetry broken solutions and is, hence, more general than the one by Bauer \textit{et al.} \cite{lit:Bauer2009}.

We obtain a rich phase-diagram at half filling and demonstrate that depending on $J$ and $U$ superconductivity, CDWs, SDWs, Kondo screening, or a different combination of these effects may occur.
Contrary to recent static mean-field calculations \cite{lit:Costa2018}, we find a novel phase at half filling where CDWs and SDWs coexist.
It is shown that 
in this phase, the SDWs lift
the degeneracy between the SC state and the CDW state such that superconductivity is suppressed.
The spectral functions reveal that the system becomes a half metal in the CDW phase near the phase boundary to the N\'eel phase.
Away from half filling, CDWs are suppressed and superconductivity survives for much larger couplings $J$.
Instead of a homogeneous N\'eel state, we observe incommensurate SDWs; however, we find no evidence that superconductivity has an influence on the pattern of this SDWs.
We show that the CDWs, at half filling, as well as the superconductivity, away from half filling, 
enhance the magnetic ordering of the localized spins since the emergent gaps in the density of states (DOS) mitigate the Kondo screening.

These results resemble recent observations 
in cuprate superconductors \cite{lit:Gabovich2010,lit:Tu2016,lit:Jang2016}.
There one can also find a rich phase diagram where superconductivity, CDWs, and SDWs coexist or compete with each other.
Similar to our model, the appearance of CDWs also strongly depends on the doping of the system.
Note, however, that cuprate superconductors are usually considered to be $d$-wave superconductors with a nonlocal pairing mechanism, 
while in this paper we only consider a local pairing.

The rest of the paper is organized as follows.
The RDMFT approach and its generalization to Nambu space are described in Sec. \ref{sec:ModelMethod}.
Furthermore, the new self-consistent NRG scheme, which is used to solve the effective impurity models, is explained in detail.
In Sec. \ref{sec:Phase diagram_half_filling}, we present the results for half filling
while
the properties of the system away from half filling are described in Sec. \ref{sec:away_from_half-filling}.
We give a short conclusion in Sec. \ref{sec:Conclusion}.

\section{Method}
\label{sec:ModelMethod}

\subsection{RDMFT setup in Nambu space}

To solve the model of Eq. \eqref{eq:lattice_H}, we employ the RDMFT, which is an extension of the conventional DMFT \cite{lit:Metzner89,lit:DMFTreview} 
to inhomogeneous situations \cite{lit:Potthoff1999}. 
It is based on the assumption of a local self-energy matrix $\underline{\Sigma}_{i,j}(\omega)=\underline{\Sigma}_i(\omega) \delta_{i,j}$, with
\begin{align}
\underline{\Sigma}_i(\omega)=
 \begin{pmatrix}
	\Sigma^i_{11}(\omega) & \Sigma^i_{12}(\omega) \\
	\Sigma^i_{21}(\omega) & \Sigma^i_{22}(\omega)
 \end{pmatrix}
\end{align}
being the self-energy matrix of site $i$ in Nambu space.
Within this approximation, correlations between different sites of the cluster are not included,
but the self-energy may be different for each lattice site and allows, therefore, e.g., SDWs and CDWs.

In the RDMFT, each site $i$ in a finite cluster 
is mapped onto its own effective impurity model with an SC symmetry breaking term
\begin{align}
 H_\mathrm{Eff} =& H_\mathrm{Imp} + \sum_{\vec{k},\sigma} \epsilon_{\vec{k},\sigma} c^\dagger_{\vec{k},\sigma} c^{\phantom{\dagger}}_{\vec{k},\sigma} 
		+\sum_{\vec{k},\sigma} V_{\vec{k},\sigma} \left( c^\dagger_{\vec{k},\sigma} d^{\phantom{\dagger}}_{\sigma} + \mathrm{H.c.} \right) \nonumber \\
		 & - \sum_{\vec{k}} \Delta_{\vec{k}} \left[ c^\dagger_{\vec{k},\uparrow} c^{\dagger}_{-\vec{k},\downarrow}  
		 + c^{\phantom{\dagger}}_{-\vec{k},\downarrow} c^{\phantom{\dagger}}_{\vec{k},\uparrow} \right], \label{eq:impurity_model}
\end{align}
where
\begin{align}
 H_\mathrm{Imp} =& \sum_{\sigma} \epsilon_d n_{d,\sigma} + U n_{d,\uparrow}n_{d,\downarrow} + J \vec{S} \vec{s}_{d}, \label{eq:impurity}
\end{align}
with $\epsilon_d=\mu$, $n_{d,\sigma} = d^\dagger_{\sigma} d^{\phantom{\dagger}}_{\sigma}$, $\vec{s}_{d}=d^\dagger_{\sigma} \vec{\sigma}_{\sigma,\sigma'} d^{\phantom{\dagger}}_{\sigma'}$ and 
$d_\sigma$ being the fermionic operator of the impurity site.
The parameters $\epsilon_{\vec{k},\sigma}$, $V_{\vec{k},\sigma}$, and $\Delta_{\vec{k}}$ are those for the medium and
may be different for each site in the RDMFT cluster.
The mapping of the lattice model of Eq. \eqref{eq:lattice_H} to the impurity model of Eq. \eqref{eq:impurity_model} is achieved by calculating the 
local Green's function in Nambu space:
\begin{align}
{G}_\mathrm{loc}(z) =& \int \int  \left[ z \idmat - H_{k_x,k_y} - \Sigma(\omega)  \right]^{-1} dk_x dk_y, \label{eq:local_Green}
\end{align}
where $H_{k_x,k_y}$ is the hopping Hamiltonian of the finite RDMFT cluster and the momentum dependence arises from the periodic boundary conditions.
The medium dependent parameters of the effective impurity model for each site $i$ are then extracted from the site-diagonal Green's function matrix in Nambu space
\begin{align}
\underline{G}_{\mathrm{loc},ii}(z) =
 \begin{pmatrix}
	\langle d^{\dagger}_{\uparrow} d^{\phantom{\dagger}}_{\uparrow} \rangle_i(z) & \langle d^{\phantom{\dagger}}_{\uparrow} d^{\phantom{\dagger}}_{\downarrow} \rangle_i(z) \\
	\langle d^{\dagger}_{\downarrow} d^{\dagger}_{\uparrow} \rangle_i(z) & \langle d^{\phantom{\dagger}}_{\downarrow} d^{\dagger}_{\downarrow} \rangle_i(z) \label{eq:local_Green_i}
 \end{pmatrix},
\end{align}
which will be discussed in detail below.

For a typical DMFT calculation, one starts with
self-energies $\Sigma_i(\omega)$ for each site of the cluster 
which should break $U(1)$ gauge symmetry to obtain an SC solution.
Afterward, the local Green's function of Eq. \eqref{eq:local_Green} is computed, which is used to set up the effective impurity problems. 
Solving these impurity models yields new self-energies $\Sigma_i(\omega)$, which are again used to calculate the local Green's functions.
This procedure is repeated until a converged solution is found.

To solve the impurity models, a variety of methods 
such as quantum Monte Carlo, exact diagonalization, or NRG \cite{lit:WilsonNRG,lit:BullaReview} can be used.
We employ the NRG to compute the self-energy and local thermodynamic quantities
of the effective impurity models
since it has been proven to be a reliable tool to calculate dynamical properties such as 
real-frequency Green's functions \cite{lit:Peters2006} and self-energies \cite{lit:Bulla98} with high accuracy around the Fermi level.
The combination of NRG and DMFT has already been successfully applied to superconductivity in interacting lattice systems \cite{lit:Bauer2009,lit:Bodensiek2013,lit:Peters2015}
although only SU(2) spin symmetric systems without magnetic ordering have been treated \cite{lit:Galitski2002,lit:Yao2014}.

\subsection{Self-consistent NRG Scheme with SU(2) spin symmetry breaking and superconductivity}
To employ the DMFT, we still have to resolve how to calculate the parameters of the NRG Wilson chain, which depend
on the local Green's function of Eq. \eqref{eq:local_Green_i} at each lattice site.
Bauer \textit{et al.} \cite{lit:Bauer2009} have shown how the DMFT+NRG setup can be extended to SC symmetry breaking.
This approach, however, requires SU(2) spin symmetry for the up and down conduction band channels.

Therefore, we propose a new and different ansatz:
Instead of directly discretizing the impurity model of Eq. \eqref{eq:impurity_model},
we first perform a Bogoliubov transformation and afterward discretize the model logarithmically into intervals $I^\alpha$ with $I^+=(x_{n+1},x_n)$ and $I^-=-(x_{n},x_{n+1})$
with $x_n=D\Lambda^{-n}$, where $\Lambda>1$ is the discretization parameter of the NRG and $D$ is the half bandwidth of the conduction band.
After retaining  only the lowest Fourier component \cite{lit:BullaReview} in Eq. \eqref{eq:impurity_model}, the Bogoliubov transformed and discretized impurity model can be written as 
\begin{align}
 H_\mathrm{Eff} =& H_\mathrm{Imp} + \sum_{\sigma,n,\alpha} \xi_{\sigma,n}^\alpha  a^\dagger_{\alpha,n,\sigma}  a^{\phantom{\dagger}}_{\alpha,n,\sigma} \nonumber \\
	 &+\sum_{n,\alpha} \left( \gamma_{n,\uparrow}^{\alpha} a^\dagger_{\alpha,n,\uparrow}d_{\uparrow} 
	 + \gamma_{n,\uparrow \downarrow}^{\alpha} a^{{\dagger}}_{\alpha,n,\uparrow}d^\dagger_{\downarrow} \right. \nonumber \\
	 & \left. + \gamma_{n,\downarrow \uparrow}^{\alpha} a^{{\dagger}}_{\alpha,n,\downarrow}d^\dagger_{\uparrow} 
	 + \gamma_{n,\downarrow}^{\alpha} a^{{\dagger}}_{\alpha,n,\downarrow}d_{\downarrow} + \mathrm{H.c.} \right). \label{eq:discretized_model}
\end{align}
The advantage of Eq. \eqref{eq:discretized_model} over the direct discretization in Ref. \cite{lit:Bauer2009} is that in each interval, the up and down conduction band channels are not directly coupled
and the U(1) gauge symmetry breaking instead occurs due to the new interval-dependent hybridizations $\gamma^\alpha_{n,\uparrow \downarrow}$ and $\gamma^\alpha_{n,\downarrow \uparrow}$.
Since the conduction band channels are not directly coupled anymore, we are able to choose the bath parameters $\xi_{\uparrow,n}^\alpha$ and $\xi_{\downarrow,n}^\alpha$ independently of each other
and, afterward, adjust the hybridizations
such that they lead to the same effective action for the impurity degree of freedom as in the original model \cite{lit:Bulla97}.
As usual in the NRG \cite{lit:BullaReview}, we can, therefore, choose $\xi_{\uparrow,n}^+ = \xi_{\downarrow,n}^+ = E^+_n = E_n$ and $\xi_{\uparrow,n}^- = \xi_{\downarrow,n}^- = E^-_n = -E_n$,
where $E_n=|x_n+x_{n+1}|/2$ is the value in the middle of an interval.

The remaining parameters for each site $i$ of the finite cluster are determined from the generalized matrix hybridization function $\underline{K}(\omega)$ in Nambu space,
which can be calculated from the local impurity Green's function matrix of Eq. \eqref{eq:local_Green_i}:
\begin{align}
 \underline{K}(z) =& z \underline{\mathbb{1}} - \underline{G}_{\mathrm{loc}}(z)^{-1} - \underline{\Sigma}(z),
\end{align}
where we have omitted the site index $i$ since the procedure is the same for every site.

To calculate the remaining parameters, we demand, as usual in the DMFT, that the local hybridization function of the lattice $\underline{K}(z)$ 
and the hybridization function of the discretized model are equal:
\begin{align}
 \underline{K}(z) =& \begin{pmatrix}
                          K^{}_{11}(z) & K^{}_{12}(z) \\
                          K^{}_{21}(z) & K^{}_{22}(z)
                         \end{pmatrix} \nonumber \\
                       = \sum_{n,\alpha} \frac{1}{z - E_n^\alpha} & \begin{pmatrix}
                          \gamma^{\alpha}_{n,\uparrow} & \gamma^{\alpha}_{n,\uparrow\downarrow} \\
                          \gamma^{\alpha}_{n,\downarrow\uparrow} & \gamma^{\alpha}_{n,\downarrow}
                         \end{pmatrix}^\dagger
                         \begin{pmatrix}
                          \gamma^{\alpha}_{n,\uparrow} & \gamma^{\alpha}_{n,\uparrow\downarrow} \\
                          \gamma^{\alpha}_{n,\downarrow\uparrow} & \gamma^{\alpha}_{n,\downarrow}
                         \end{pmatrix}.                    
\end{align}
Since $K_{12}(z) = K_{21}(z)$ must apply, we can choose $\gamma^{\alpha}_{n,\downarrow\uparrow} = \gamma^{\alpha}_{n,\uparrow\downarrow}=\gamma^{\alpha}_{n,\mathrm{off}}$.
Using only the imaginary parts $\Delta_\uparrow(\omega)=-\mathrm{Im}\, K_{11}(\omega+i\eta)/\pi$, $\Delta_\downarrow(\omega)=-\mathrm{Im}\, K_{22}(\omega+i\eta)/\pi$,
and $\Delta_\mathrm{off}(\omega)=-\mathrm{Im} \, K_{12}(\omega+i\eta)/\pi$, the equation can be rewritten as a sum of delta functions
\begin{align}
 \Delta_\uparrow(\omega) =& \sum_{n,\alpha} ( {\gamma^{\alpha}_{n,\uparrow}}^2 + {\gamma^{\alpha}_{n,\mathrm{off}}}^2 ) \delta(\omega - E_n^\alpha),  \\
 \Delta_\downarrow(\omega) =& \sum_{n,\alpha} ( {\gamma^{\alpha}_{n,\downarrow}}^2 + {\gamma^{\alpha}_{n,\mathrm{off}}}^2 ) \delta(\omega - E_n^\alpha),  \\
 \Delta_\mathrm{off}(\omega) =& \sum_{n,\alpha} \gamma^\alpha_{n,\mathrm{off}} ( \gamma^{\alpha}_{n,\uparrow} + \gamma^{\alpha}_{n,\downarrow} ) \delta(\omega - E_n^\alpha).
\end{align}
Integration over the energy intervals $I^\alpha_n$,
\begin{align}
 w^\alpha_{n,\sigma} = \int_{I^\alpha_n} \Delta_{\sigma}(\omega) d\omega \quad
 w^\alpha_{n,\mathrm{off}} = \int_{I^\alpha_n} \Delta_{\mathrm{off}}(\omega) d\omega,
\end{align}
yields the equation system
\begin{align}
 w^\alpha_{n,\uparrow}     =& {\gamma^{\alpha}_{n,\uparrow}}^2 + {\gamma^{\alpha}_{n,\mathrm{off}}}^2, \\
 w^\alpha_{n,\downarrow}   =& {\gamma^{\alpha}_{n,\downarrow}}^2 + {\gamma^{\alpha}_{n,\mathrm{off}}}^2, \\
 w^\alpha_{n,\mathrm{off}} =& \gamma^\alpha_{n,\mathrm{off}} ( \gamma^{\alpha}_{n,\uparrow} + \gamma^{\alpha}_{n,\downarrow} ).
\end{align}
One possible solution of this system is given by
\begin{align}
 \gamma^{\alpha}_{n,\uparrow} =& \frac{ w^\alpha_{n,\uparrow} + \sqrt{ w^\alpha_{n,\uparrow}  w^\alpha_{n,\downarrow} - {w^\alpha_{n,\mathrm{off}}}^2 } }{
\sqrt{w^\alpha_{n,\uparrow} + w^\alpha_{n,\downarrow} +2\sqrt{w^\alpha_{n,\uparrow} w^\alpha_{n,\downarrow} -  {w^\alpha_{n,\mathrm{off}}}^2} } }, \\ 
\gamma^{\alpha}_{n,\downarrow} =& \frac{ w^\alpha_{n,\downarrow} + \sqrt{ w^\alpha_{n,\uparrow}  w^\alpha_{n,\downarrow} - {w^\alpha_{n,\mathrm{off}}}^2 } }{
\sqrt{w^\alpha_{n,\uparrow} + w^\alpha_{n,\downarrow} +2\sqrt{w^\alpha_{n,\uparrow} w^\alpha_{n,\downarrow} -  {w^\alpha_{n,\mathrm{off}}}^2} } }, \\ 
\gamma^{\alpha}_{n,\mathrm{off}} =& \frac{ w^\alpha_{n,\mathrm{off}}}{
\sqrt{w^\alpha_{n,\uparrow} + w^\alpha_{n,\downarrow} +2\sqrt{w^\alpha_{n,\uparrow} w^\alpha_{n,\downarrow} -  {w^\alpha_{n,\mathrm{off}}}^2} } }.
\end{align}
Note that in the case of vanishing superconductivity $w^\alpha_{n,\mathrm{off}}=0$, the equations reduce to the standard NRG solution \cite{lit:BullaReview}
${\gamma^{\alpha}_{n,\sigma}}^2= w^\alpha_{n,\sigma}$ and $\gamma^{\alpha}_{n,\mathrm{off}}=0$.

\begin{figure}[t]
	\includegraphics[width=0.49\textwidth]{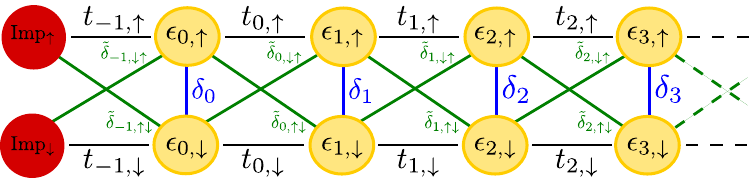}
	\caption{
	New Wilson chain with the superconducting symmetry breaking terms $\delta_{n}$ (blue lines), $\tilde\delta_{n,\downarrow\uparrow}$ and $\tilde\delta_{n,\uparrow\downarrow}$ (green lines).
	$\tilde\delta_{n,\downarrow\uparrow}$ and $\tilde\delta_{n,\uparrow\downarrow}$ vanish in the case of SU(2) spin symmetry.
	}
	\label{fig:WilsonChain}
\end{figure}
Now that we have calculated all model parameters from a given hybridization function $\underline{K}(\omega)$,
the next step is to map the impurity model of Eq. \eqref{eq:impurity_model} via a Householder transformation to a linear chain model of the form
\begin{align}
 H_\mathrm{Eff} =& H_\mathrm{Imp} + \sum_{n=0,\sigma}^N \epsilon_{n,\sigma} f^\dagger_{n,\sigma} f^{\phantom{\dagger}}_{n,\sigma} 
		  + \sum_{n=0}^N \delta_{n} \left(f^\dagger_{n,\uparrow} f^{{\dagger}}_{n,\downarrow} + \mathrm{H.c.} \right) \nonumber \\ 
		 &+ \sum_{n=-1}^{N-1} \left( \tilde\delta_{n,\uparrow\downarrow} f^\dagger_{n,\uparrow} f^{{\dagger}}_{n+1,\downarrow}
		  + \tilde\delta_{n,\downarrow\uparrow} f^\dagger_{n,\downarrow} f^{{\dagger}}_{n+1,\uparrow} + \mathrm{H.c.} \right) \nonumber \\
		  & + \sum_{n=-1,\sigma}^{N-1} t_{n,\sigma} \left( f^\dagger_{n,\sigma} f^{\phantom{\dagger}}_{n+1,\sigma} + \mathrm{H.c.} \right).
\end{align}
The new Wilson chain is illustrated in Fig. \ref{fig:WilsonChain}.
In addition to the usual hopping parameters $t_{n,\sigma}$ and on-site energies $\epsilon_{n,\sigma}$ of an ordinary Wilson chain,
this chain exhibits the SC symmetry breaking terms $\delta_{n}$ (blue lines), $\tilde\delta_{n,\downarrow\uparrow}$ and $\tilde\delta_{n,\uparrow\downarrow}$ (green lines).
In the case of SU(2) spin symmetry the terms $\tilde\delta_{n,\downarrow\uparrow}$ and $\tilde\delta_{n,\uparrow\downarrow}$ vanish 
and the chain reduces to the form of Bauer \textit{et al.} \cite{lit:Bauer2009}.

Since $\tilde \delta_{n,\downarrow\uparrow}$ and $\tilde \delta_{n,\uparrow\downarrow}$ link different energy scales, 
it is important to emphasize that both terms decay exponentially with increasing $n$ and, therefore, ensure the separation of energy scales which is vital for the NRG.
Also note that both terms do not need to be equal but depend on the details of the Householder transformation, e.g. it is also possible that one of these terms always vanishes.

Since the described NRG scheme is completely independent of $H_\mathrm{Imp}$, which incorporates all impurity degrees of freedom,
we have tested it for the exactly solvable case of vanishing Hubbard $U=0$ and Kondo coupling $J=0$
and found good agreement.

\section{Half Filling}
  \label{sec:Phase diagram_half_filling}
  
  \subsection{Phase diagram }

\begin{figure}[t]
	\includegraphics[width=0.49\textwidth]{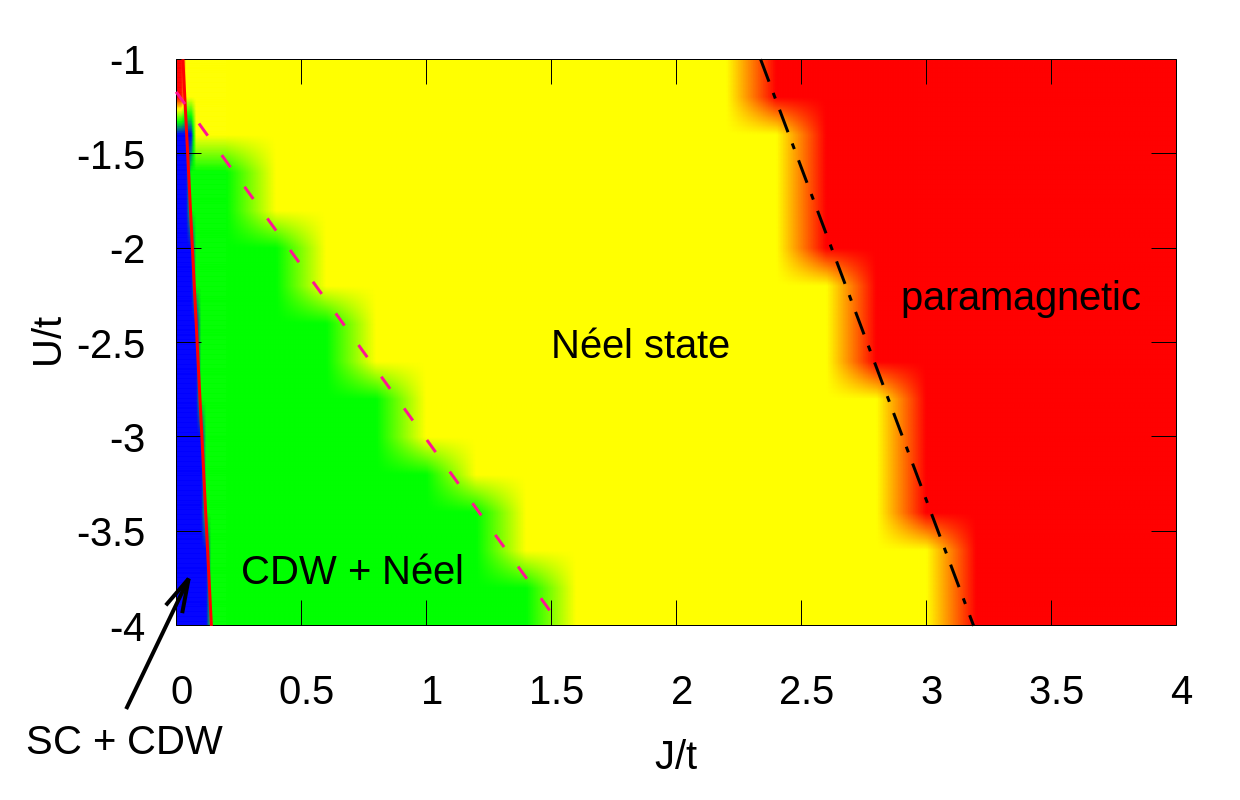}
	\caption{
	    The phase diagram at half filling in dependence of an attractive $U$ and antiferromagnetic Kondo coupling $J$.
	    A detailed explanation is given in the text.
	    Lines indicate phase boundaries (see Fig. \ref{fig:Critical_J}).
	    The step structures around the phase boundaries are caused by the finite resolution of the data.
	}
	\label{fig:phase_diagram}
\end{figure}
\begin{figure}[t]
	\flushleft{(a)}
	\includegraphics[width=0.49\textwidth]{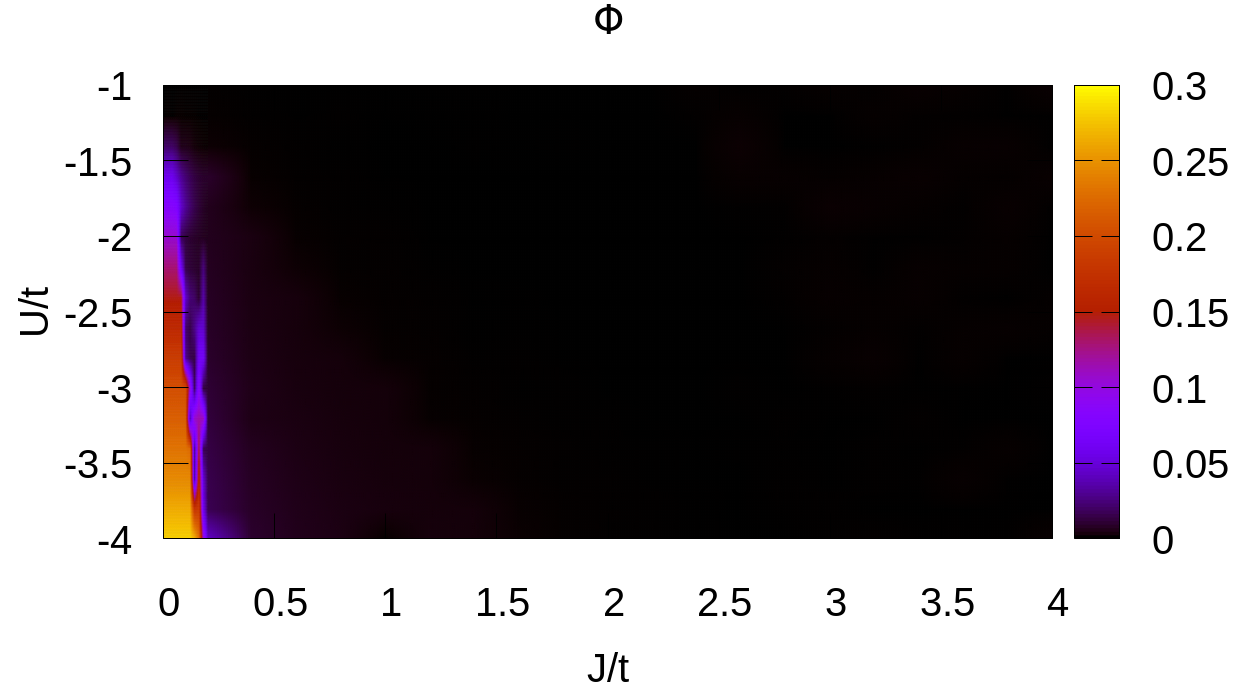}
	\flushleft{(b)}
	\includegraphics[width=0.49\textwidth]{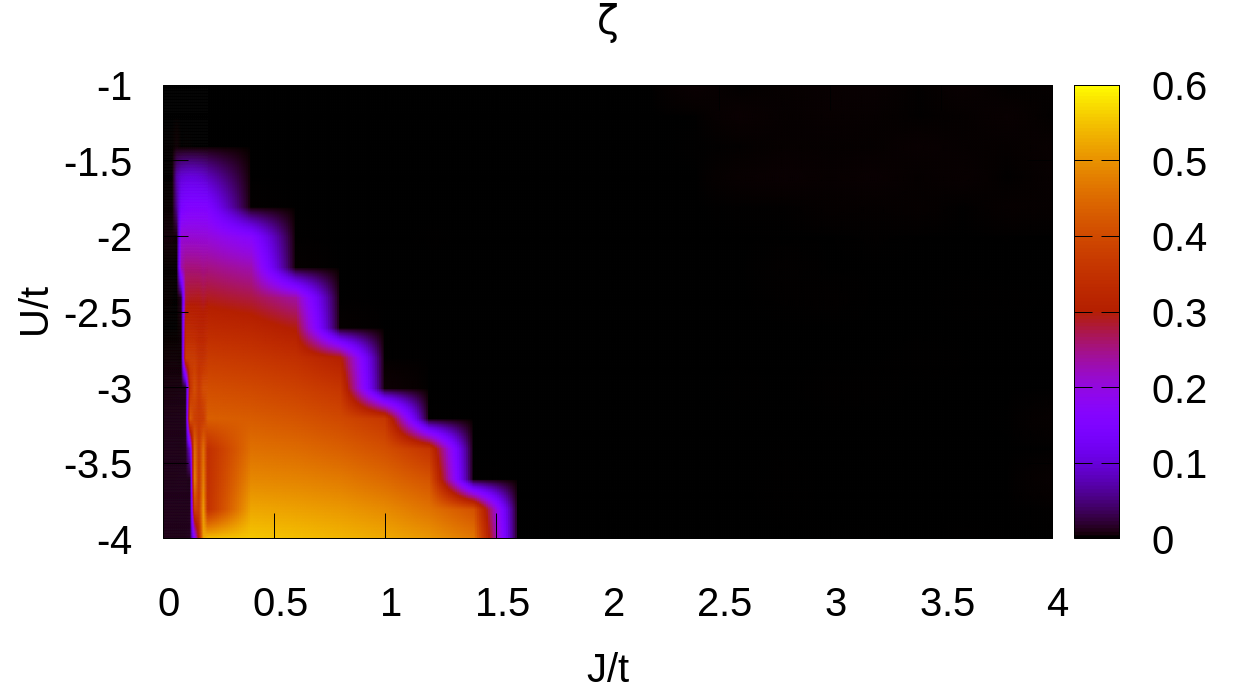}
	\flushleft{(c)}
	\includegraphics[width=0.49\textwidth]{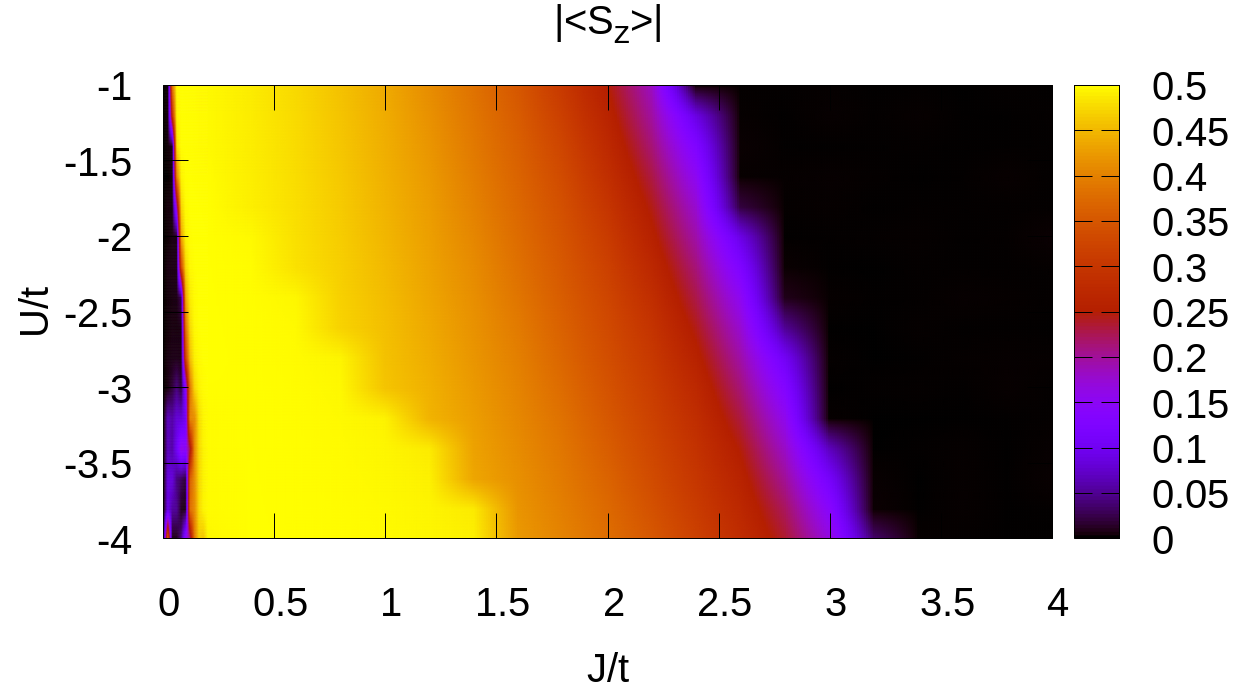}
	\caption{
	    Different order parameters of the system at half filling as a function of $U$ and $J$:
	    (a) The anomalous expectation value $\Phi=\langle d^\dagger_\uparrow d^\dagger_\downarrow \rangle$.
	    (b) The CDW order parameter $\zeta=|n_{d,i}-n_{d,i+1}|/2$ .
	    (c) The polarization of the localized $f$-electron spins $|\langle S_z \rangle|$.
	    The step structures around the phase boundaries are caused by the finite resolution of the data.
	}
	\label{fig:phase_diagram_properties}
\end{figure}
\begin{figure}[t]
	\includegraphics[width=0.49\textwidth]{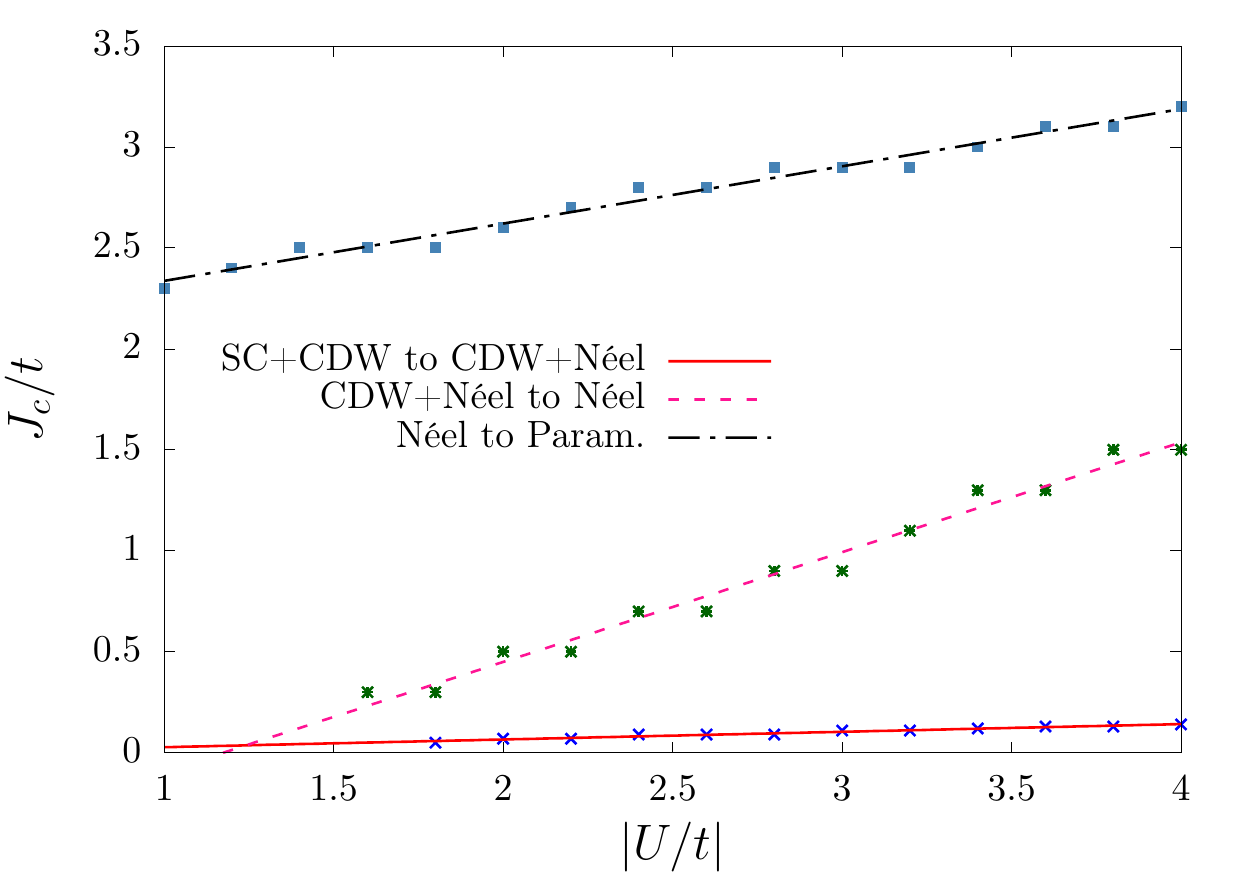}
	\caption{
	    Critical couplings $J_c(U)$ separating the phases plotted against $|U|$.
	    We find a linear behavior for all three phase boundaries:
	    $J_c/t = 0.038 |U/t| - 0.011$ for the transition from SC+CDW to CDW+N\'eel (red solid line),
	    $J_c/t = 0.544 |U/t| - 0.638$ for the transition from CDW+N\'eel to N\'eel (pink dashed line), and
	    $J_c/t = 0.284 |U/t| + 2.053$ for the transition from N\'eel to paramagnetism (black dashed-dotted line).
	}
	\label{fig:Critical_J}
\end{figure}
Figure \ref{fig:phase_diagram} summarizes our main results and depicts the phase diagram as a function of the strength of the attractive $U$ and antiferromagnetic Kondo coupling $J$ for half filling.
The calculations are performed for $T/t=4\cdot10^{-5}$.

For a vanishing coupling $J$, our observations are in agreement with the previous results for an attractive Hubbard model \cite{lit:Huscroft1997,lit:Bauer2009}.
At half filling and $J=0$, the SC state is energetically degenerate with a CDW state so that an arbitrary superposition of both states yields a stable solution in the DMFT.
For a CDW state, the occupation of each lattice site may differ from half filling, but the average of two neighboring sites yields $n_d=1$, with $n_d=n_{d,\uparrow} + n_{d,\downarrow}$,
such that on average the whole lattice is half filled.

This behavior does not change for very weak couplings $J$; namely, we also see SC solutions for finite couplings.
This is demonstrated in Fig. \ref{fig:phase_diagram_properties}(a), 
which shows the anomalous expectation value $\Phi=\langle d^\dagger_\uparrow d^\dagger_\downarrow \rangle$ as a color contour plot.
In this regime, the system behaves exactly as in the $J=0$ case and we do not observe magnetic ordering for the localized spins since the Ruderman-Kittel-Kasuya-Yosida (RKKY) interaction is too weak.
Notice that for these small coupling strengths $J$, the spins and the conduction electrons are effectively decoupled at this temperature.

For larger couplings, the system undergoes a first-order transition.
The superconductivity breaks down and the anomalous expectation value exhibits a jump to $\Phi\approx 0$.
The small residual value of $\Phi$ might be caused by a finite spectral resolution due to numerical noise and broadening of the NRG spectra.
Note, however, that a finite temperature in a real experiment would have a similar effect and could also lead to a finite SC expectation value $\Phi$.
The reason for this is that the energy difference between the SC state and the CDW state is very small
so that the SC state is partially occupied due to the finite temperature effect.

For these larger coupling strengths $J$, the CDW phase is energetically favoured over the superconductivity without the need of a nonlocal density-density interaction. 
We thus observe a CDW phase, which is revealed in Fig. \ref{fig:phase_diagram_properties}(b) that depicts the CDW order parameter $\zeta=|n_{d,i}-n_{d,i+1}|/2$,
measuring the difference in the occupation of two neighboring sites.
Figure  \ref{fig:phase_diagram_properties}(c) displays the polarization of the localized $f$-electron spins.
In addition to the onset of the CDWs, we also find SDWs where the localized spins are ordered in an antiferromagnetic N\'eel state.
The bright yellow area in Fig. \ref{fig:phase_diagram_properties}(c) indicates that in this regime the spins are almost completely polarized
since the Kondo screening is suppressed due to the relatively large gap created by the CDW at the Fermi energy in the DOS.

Note that although the degeneracy is lifted in this phase,
the energy difference between CDW and superconductivity is very small such that it may take a large number of DMFT iterations 
to go from an SC solution to a CDW solution.
The critical coupling separating the two phases displays a linear dependence on $U$, as depicted in Fig. \ref{fig:Critical_J} (red solid line).
However, the gradient is very small such that the critical couplings are very similar for a wide range of $U$.

The reason why the CDW state has lower energy compared to the SC state is
that the antiferromagnetically ordered $f$-electron spins generate magnetic fields which oscillate from site to site.
In an SC state, a magnetic field always decreases the gap size while
in a CDW state it is possible to retain the size of the gap in at least one of the conduction band channels, i.e., up- or down-spin channel.
Consequently, the system becomes a half metal in this phase since the gap closes only in one of the conduction band channels. 
Thus, antiferromagnetically ordered spins can cooperate with a CDW order in the conduction electrons, but not with SC conduction electrons.
This will be discussed in more detail in Sec. \ref{sec:Dynamic_properties}.

We point out that this CDW+N\'eel phase and the breakdown of superconductivity has not been observed in recent static mean-field theory calculations \cite{lit:Costa2018}.
Instead, a phase combining SDWs and superconductivity has been found
because CDWs have not been considered in this static mean field approach 
while CDWs emerge in our RDMFT framework without any additional assumptions.

Upon further increasing the Kondo coupling, another first-order transition, indicated by discontinuous jumps in physical properties, is observed and the CDW vanishes.
The critical coupling shows again a linear dependence on $U$ as indicated by the dashed pink line in Fig. \ref{fig:Critical_J}.
Note that in this phase, the polarization of the localized spins decreases [see Fig. \ref{fig:phase_diagram_properties}(c)]. 
The reason for this is a change in the size of the gap in the DOS at the transition from the CDW to the N\'eel phase.
The Kondo temperature 
$T_k=D\mathrm{e}^{-1/\rho J}$ exponentially depends on the coupling $J$ and the DOS around the Fermi energy $\rho$.
In the CDW phase, the gap is rather large, which impedes the Kondo effect, while in the N\'eel phase the gap in the DOS becomes significantly smaller.
This leads to an increased Kondo screening in the N\'eel phase and, hence, a decrease of the spin polarization.

For larger couplings, the Kondo temperature exponentially increases and
we obtain the results of a standard Kondo lattice model without an additional attractive interaction, $U=0$ \cite{lit:Peters2015_2}.
Close to half filling,
the Kondo lattice is dominated by the interplay 
between RKKY interaction $\propto J^2$ and the Kondo effect as described by the Doniach phase diagram \cite{lit:Doniach77}.
For relatively small couplings $J$, the localized $f$-electrons are antiferromagnetically ordered in a N\'eel state, thus, suppressing the Kondo effect.
On the other hand, with increasing coupling the Kondo screening becomes more dominant such that the polarization of the localized spins decreases.

At strong couplings, the Kondo effect dominates and the system undergoes a continuous transition from a magnetically ordered N\'eel state to a paramagnetic state \cite{lit:Peters2015_2}.
Compared to the $U=0$ case, the critical coupling at which the transition from the magnetically ordered to the paramagnetic state occurs, increases for a finite attractive $U$.
Again, a linear dependence on $U$ is found for the critical Kondo coupling, which is depicted in Fig. \ref{fig:Critical_J} as a black dashed-dotted line.
Note that the constant offset of about $2.053$, which indicates the critical coupling $J_c$ for the case of vanishing interaction $U=0$,
is in good agreement with the results of a standard Kondo lattice without additional attractive interactions \cite{lit:Otsuki2015,lit:Peters2015_2}.
The reason for the increasing critical coupling is that with increasing attractive $U$, either the doubly occupied or empty state with total spin $s=0$
is favored over the singly occupied state with $s=1/2$ and,
consequently, the effective magnetic moment in the conduction band, which screens the localized spins, vanishes.
Therefore, an attractive interaction $U$ inhibits Kondo screening of the localized spins \cite{lit:Raczkowski2010}.

\subsection{Static properties and phase transitions}
\label{sec:Static_properties}
\begin{figure}[t]
	\includegraphics[width=0.49\textwidth]{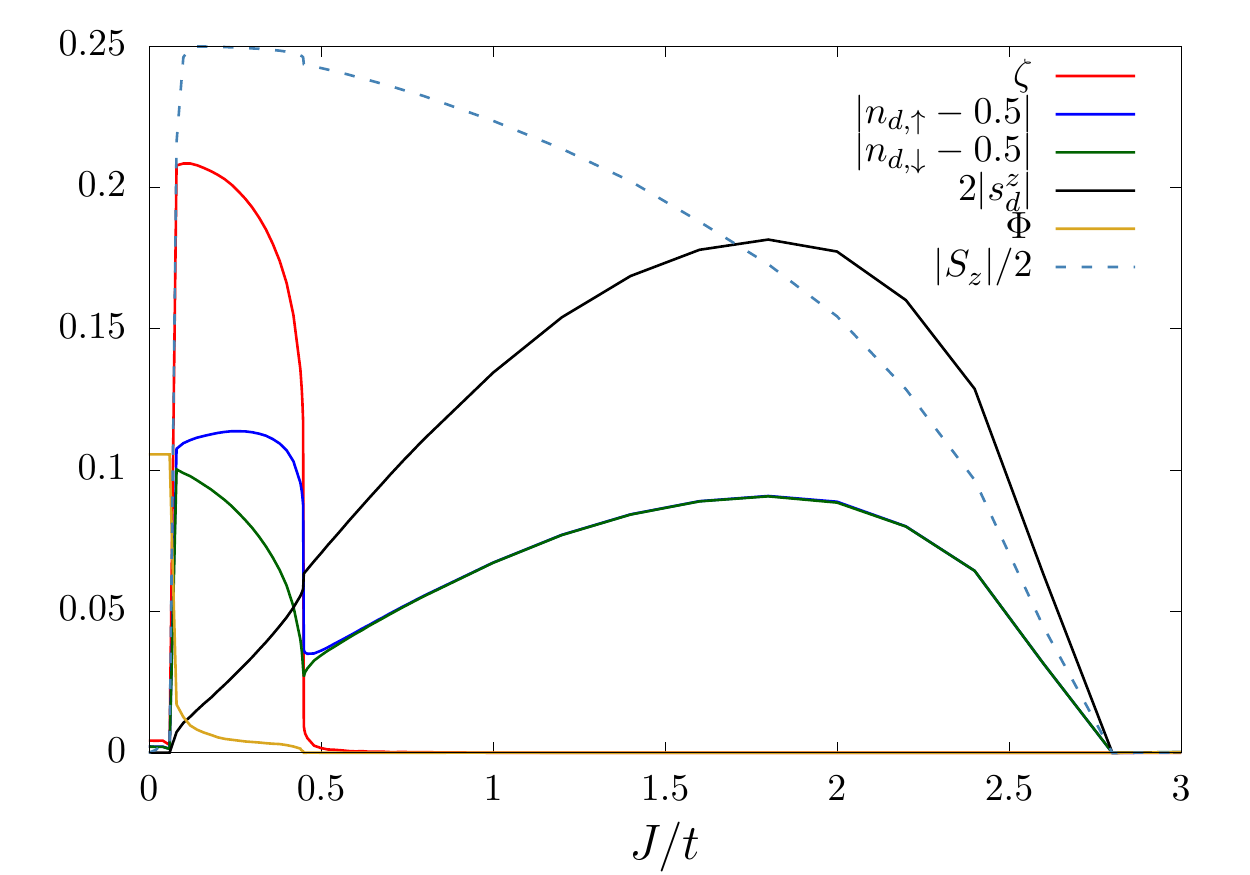}
	\caption{
		Occupation of both spin-channels $|n_{d,\sigma}-0.5|$, CDW order parameter $\zeta$,
		polarization of the conduction band $2|s^z_d|=|n_{d,\uparrow} - n_{d,\downarrow}|$, anomalous expectation value $\Phi$,
		and $f$-electron spin polarization $|S_z|$ as functions of the coupling $J$
		for a constant $U/t=-2$ exactly at half filling.
	}
	\label{fig:U-0100_JScan_half-filling}
\end{figure}
We now discuss the static properties of the system in greater detail.
Figure \ref{fig:U-0100_JScan_half-filling} shows the deviation of the occupation numbers $n_{d,\uparrow}$ and $n_{d,\downarrow}$ from half filling, the CDW order parameter $\zeta$,
the anomalous expectation value $\Phi=\langle d^\dagger_\uparrow d^\dagger_\downarrow \rangle$, and the spin polarization of the localized $f$-electrons
for a constant attractive $U/t=-2$ as a function of the coupling $J$. 

For small couplings up to $J/t\approx 0.065$, the system behaves exactly in the same way as for $J=0$; 
the SC and CDW states are degenerate (CDW state not explicitly shown).
The anomalous expectation value is constant and since we start with a non-CDW self-energy, the occupation for all sites is exactly half filling $|n_{d,\sigma}-0.5|=0$ and $\zeta=0$.
Due to the small coupling, the RKKY interacting is very weak and we do not observe a magnetic ordering of the localized $f$-electron spins.
Note, however, that the localized spins are completely unscreened, due to the SC gap, see Sec. \ref{sec:Dynamic_properties} below.
Therefore, the localized $f$-electrons essentially behave like free spins and
even very weak perturbations can polarize them.
In the current model, however, there is no possibility to mediate the coupling between spins other than the RKKY interaction.
In a real material, it is very likely that different long range interactions would produce a stable magnetic ordering in this phase.

For larger couplings $J$, all properties 
show a discontinuous jump indicating a first-order transition.
The degeneracy between superconductivity and CDWs is lifted so that superconductivity almost completely vanishes and instead a CDW state with $\zeta \neq 0$ appears.
Upon further increasing the coupling, the conduction-band polarization $2|s^z_d|=|n_{d,\uparrow} - n_{d,\downarrow}|$ continuously increases due to the magnetic fields induced by the localized spins
and, consequently, the small residual superconductivity eventually vanishes.
Note, however, that only the occupation $n_{d,\downarrow}$ changes while $n_{d,\uparrow}$ remains almost constant.
The reason for this behavior is the alternating magnetic fields originating from the antiferromagnetic ordered $f$-electrons, 
which enable the system to preserve the gap in at least one spin-channel.

Around $J/t\approx 0.45$, another first-order transition occurs and most physical properties display a discontinuous jump.
The occupation number at each lattice site jumps to half filling such that $\zeta=0$ and $|n_{d,\uparrow}-0.5|=|n_{d,\downarrow}-0.5|$.
For the localized spins, we find the typical N\'eel state of a Kondo lattice model at half filling without any CDW or SC order $\Phi=0$.
Note the small jump in $S_z$ ($s_d^z$) at the phase transition around $J/t\approx 0.5$, indicating that the polarization is slightly smaller (larger) than the one in the CDW phase.

For even larger couplings, the well known second-order transition for the standard Kondo lattice from a magnetically ordered to a paramagnetic state occurs \cite{lit:Peters2015_2}
and the N\'eel state vanishes continuously.

\subsection{Dynamical properties}
  \label{sec:Dynamic_properties}
\begin{figure}[t]
	\flushleft{(a)}
	\includegraphics[width=0.49\textwidth]{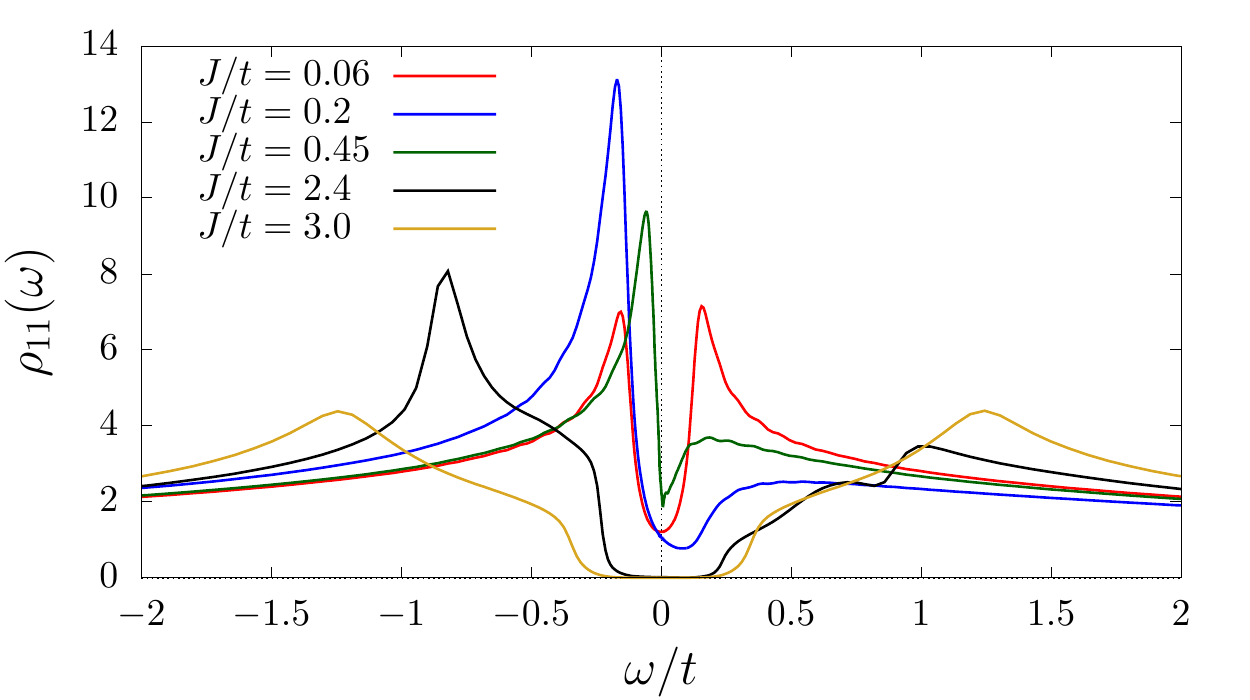}
	\flushleft{(b)}
	\includegraphics[width=0.49\textwidth]{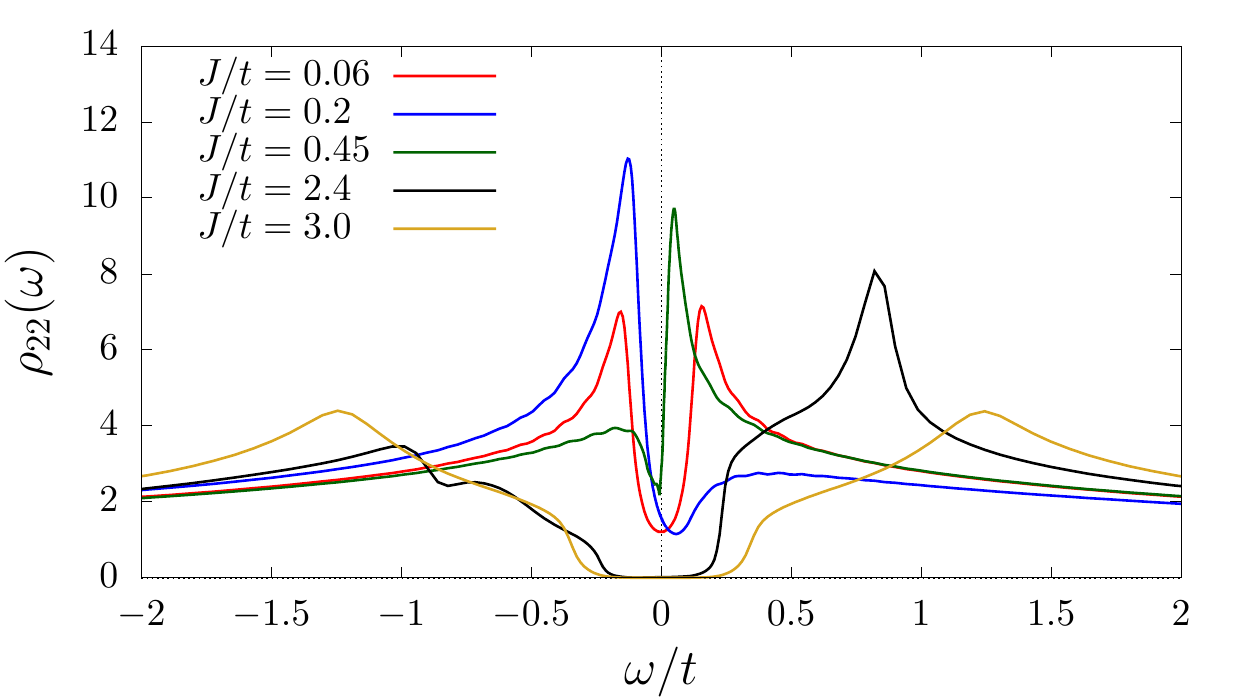}
	\flushleft{(c)}
	\includegraphics[width=0.49\textwidth]{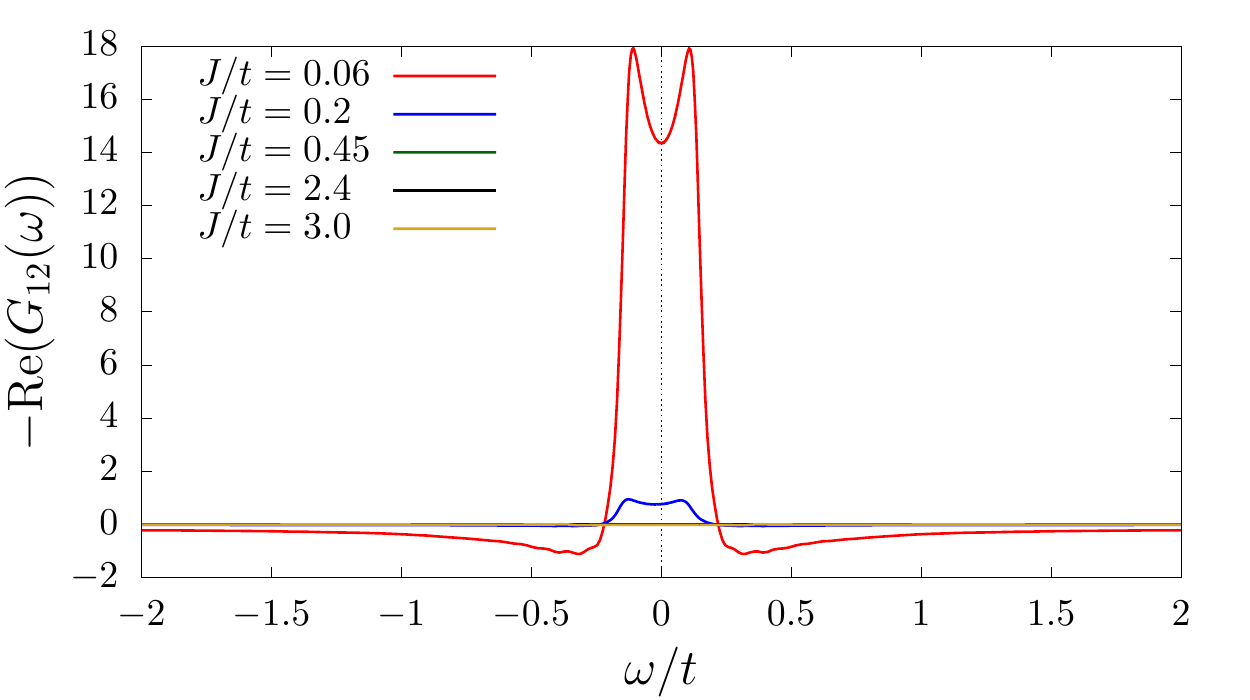}
	\caption{
	    The local DOS (a) $\rho_{11}(\omega)$ for the spin-up and (b) $\rho_{22}(\omega)$ for the spin-down channel for $U/t=-2$.
	    The spectrum at the neighboring sites is mirrored on the $\omega=0$ axis.
	    (c) Real part of the off-diagonal Green's function.
	    The system is for $J/t=0.06$ in the SC, for $J/t=0.2$ in the CDW, for $J/t=0.45$, and $J/t=2.4$ in the N\'eel and for $J/t=3.0$ in the paramagnetic phase.
	}
	\label{fig:Local_Greens_function}
\end{figure}

The local DOSs at a lattice site for the spin-up and spin-down channel are depicted in Figs. \ref{fig:Local_Greens_function}(a) and \ref{fig:Local_Greens_function}(b), respectively,
for different couplings $J$ and constant $U/t=-2$.
The spectrum at neighboring sites is mirrored on the $\omega=0$ axis.
Figure \ref{fig:Local_Greens_function}(c) displays the real part of the off-diagonal Green's function, where finite values indicate superconductivity.

For the very weak coupling $J/t=0.06$ (red line), we observe a gap with two symmetric peaks for both spin channels.
Since the magnetic order is absent for this small coupling, the spectrum for the spin-up and spin-down channel is identical.
The pronounced value of the $\mathrm{Re}[G_{12}(\omega)]$ inside the gap shows that this gap originates from superconductivity.

In the CDW phase for $J/t=0.2$ (blue line), both spin channels have just one peak below the Fermi energy, which is shifted to energies above the Fermi energy
for neighboring sites due to the CDW.
Note that the position and the heights of the peaks for the two channels is not identical.
The value of the off-diagonal Green's function is very small, in accordance with the observation of a very small $\Phi$, see Fig. \ref{fig:U-0100_JScan_half-filling}.
In this phase, the localized spins are almost completely polarized since the gap at the Fermi energy suppresses the screening of the $f$-electrons due to the Kondo effect.

\begin{figure}[t]
	\flushleft{(a)}
	\includegraphics[width=0.49\textwidth]{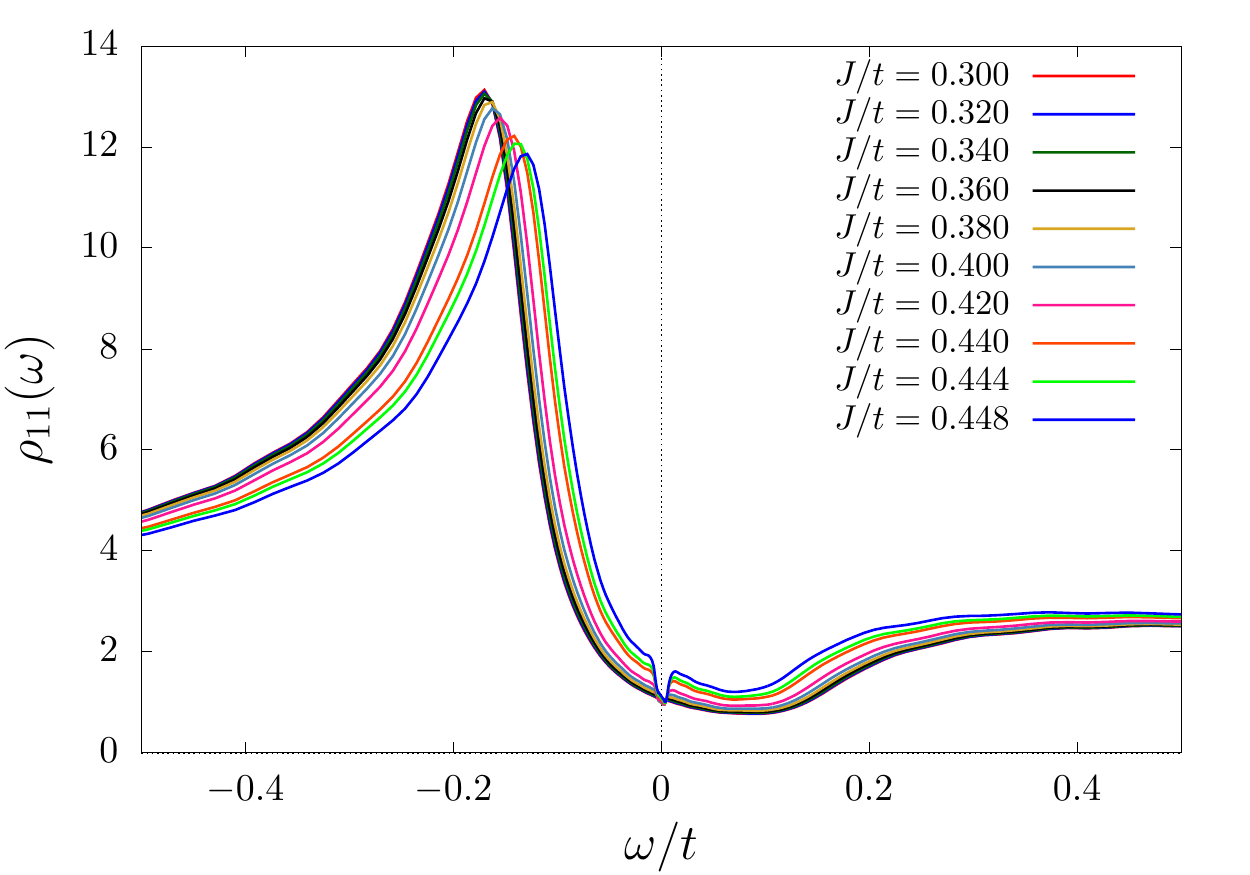}
	\flushleft{(b)}
	\includegraphics[width=0.49\textwidth]{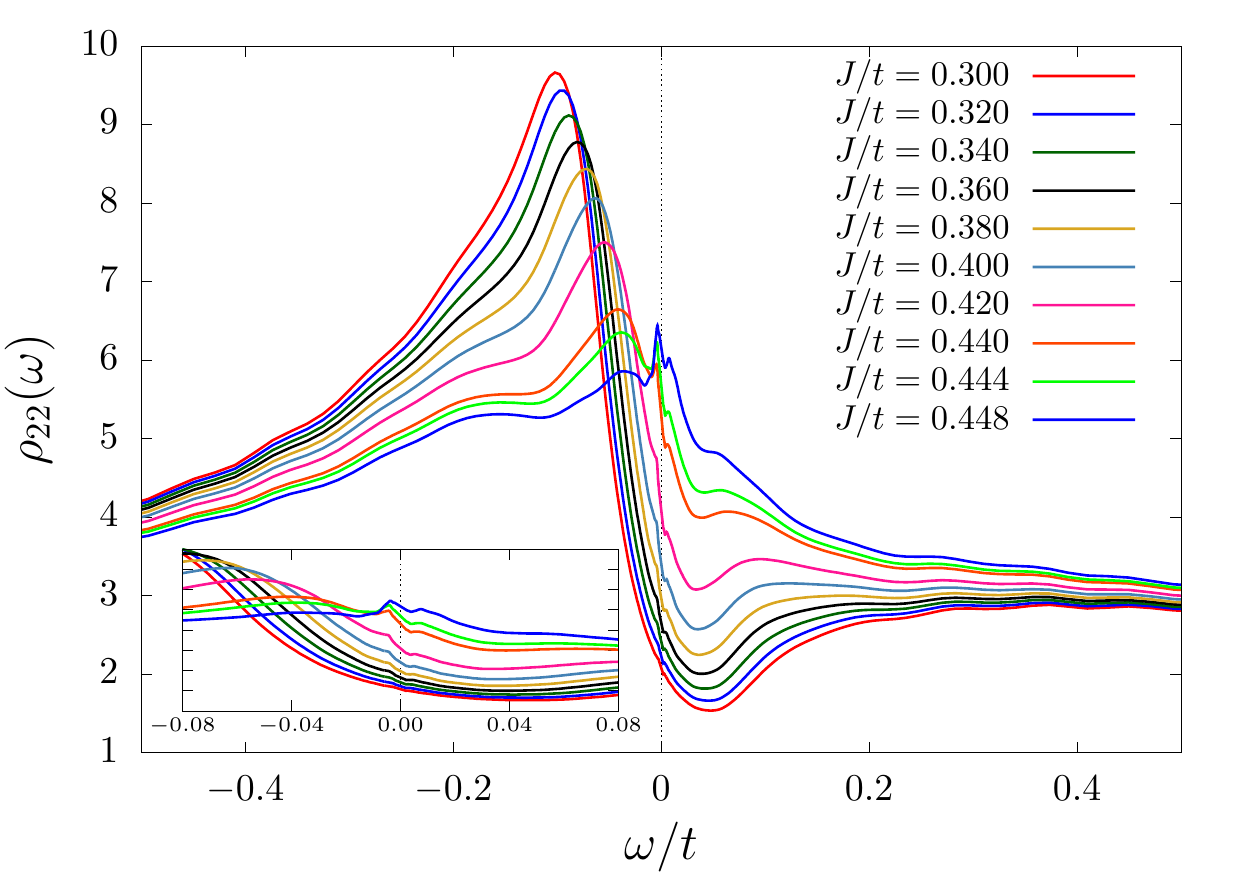}
	\caption{
	    Local DOS (a) $\rho_{11}(\omega)$ for the spin-up and (b) $\rho_{22}(\omega)$ for the spin-down channel 
	    in the CDW + N\'eel state for different couplings close to the phase transition to the magnetically ordered phase and $U/t=-2$.
	    For neighboring sites, the spectrum is mirrored on the $\omega=0$ axis.
	}
	\label{fig:Local_Greens_function_CDW}
\end{figure}

For the two larger couplings $J/t=0.45$ (green line) and $J/t=2.4$ (black line),
the system is in the magnetically ordered N\'eel phase already known for the ordinary Kondo lattice.
The off-diagonal Green's function completely vanishes, showing that there is no superconductivity anymore in this phase.
In this phase, the size of the gap strongly depends on the strength of the coupling $J$ \cite{lit:Coleman2006}
so that the gap is very small for $J/t=0.45$ while it is rather large for $J/t=2.4$.
The small gap for weak couplings $J$ in the N\'eel phase leads to a sudden increase of the DOS around the Fermi energy compared to the CDW phase and, consequently, 
results in an enhancement of the Kondo effect.
The increasing influence of the Kondo effect causes a decrease of the polarization of the localized spins, 
which can be seen as a small jump in $S_z$ at the phase transition [see, e.g., Fig. \ref{fig:phase_diagram_properties}(c)].

At the phase transition point from the CDW to the N\'eel phase, the peak in $\rho_{22}(\omega)$ discontinuously jumps from below to above the Fermi energy,
once again indicating a first-order transition.

For the coupling $J/t=3.0$, the system is in the paramagnetic phase and we observe a gap with two symmetric peaks.
Since there is no polarization anymore, the DOS of the spin-up and spin-down channel are identical.
As before, we do not observe superconductivity and $\mathrm{Re}[G_{12}(\omega)]$ is completely zero.

Since, for very small couplings, the system behaves just like an attractive Hubbard model with $J=0$ while, for large couplings, the results of a standard Kondo lattice with $U=0$ are obtained,
the CDW phase is the most interesting phase.
Figure \ref{fig:Local_Greens_function_CDW}, therefore, depicts the local DOS in the CDW phase for couplings close to the phase transition point to the magnetically ordered state in more detail.
When approaching the phase transition point to the SDW phase, we observe that the position of the peak for the spin-up channel is almost unchanged, indicating an insulating system. 
On the other hand, the peak in the spin-down channel is shifted towards the Fermi energy, leading to a gap-closing.
In addition to the shift, the spectral weight at the Fermi energy increases and two small peaks around the $\omega=0$ evolve.
The energy scale on which these additional peaks appear agrees very well with the energy of $J \langle \vec{S}\cdot\vec{s}_{d}\rangle$, 
indicating that these peaks originate from spin-flip excitations.

A comparison between Figs. \ref{fig:Local_Greens_function_CDW}(a) and \ref{fig:Local_Greens_function_CDW}(b) reveals that the system becomes a half metal close to the quantum phase transition where only the gap in one conduction band channel disappears.
This behavior arises from the combination of the CDW and the oscillating magnetic fields caused by the localized spins.
While at the site shown in Fig. \ref{fig:Local_Greens_function_CDW} the effective magnetic field tends to shift the peak in $\rho_{11}(\omega)$ to lower energies and away from the Fermi energy, 
the peak in $\rho_{22}(\omega)$ is displaced toward the Fermi energy.
At the neighboring sites, the situation is the same.
Because of the spin-flip of the localized spin in the N\'eel state, the effective magnetic field now shifts the peak of $\rho_{11}(\omega)$ to higher energies. 
However, due to the CDW, the peak is now located above $\omega=0$ so that it is again displaced away from the Fermi energy.
For the same reason, the peak in $\rho_{22}(\omega)$ of the neighboring sites is shifted toward the Fermi energy so that the gap closes only for the spin-down channel.

\begin{figure}[t]
	\flushleft{(a)}
	\includegraphics[width=0.49\textwidth]{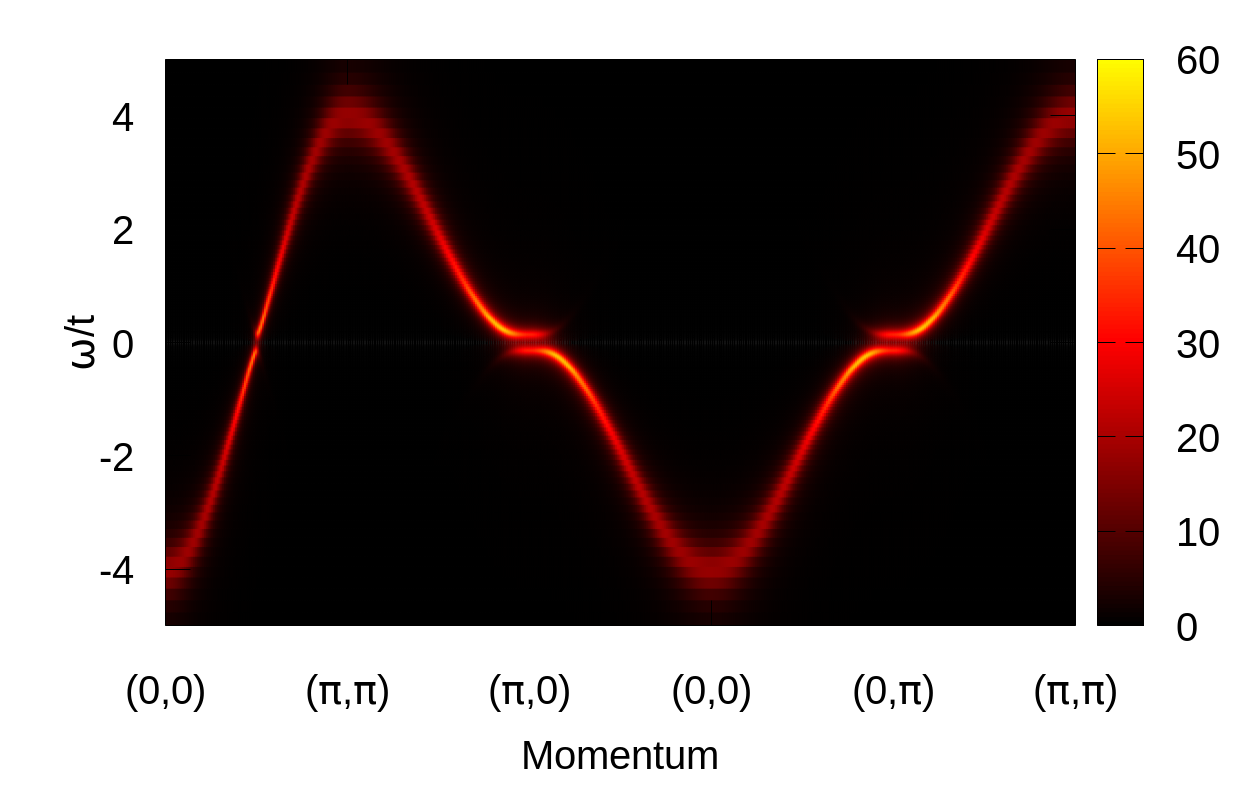}
	\flushleft{(b)}
	\includegraphics[width=0.49\textwidth]{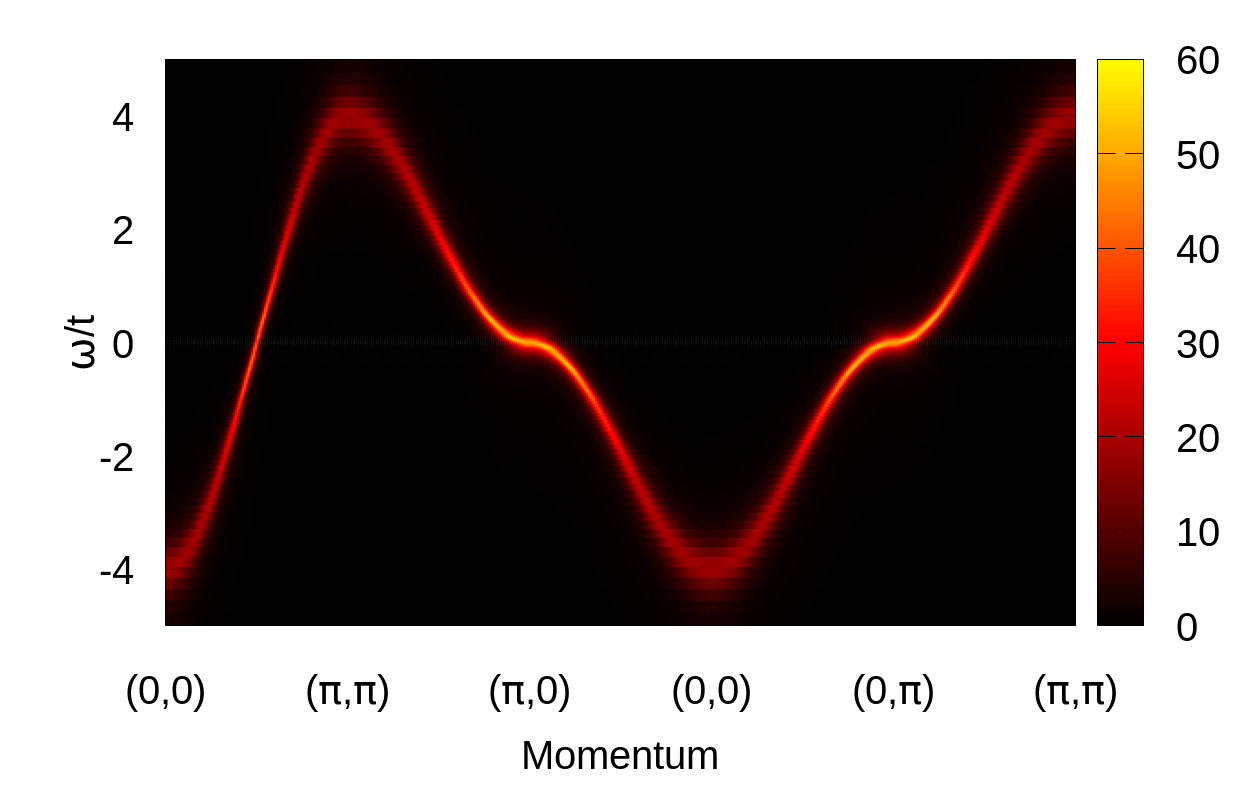}
	\caption{
		Momentum-dependent spectral functions of (a) the spin-up channel and (b) spin-down channel 
		for $U/t=-2$ and a coupling $J/t=0.4$ close to the quantum phase transition.
		}
	\label{fig:Momentum_Spectral_function}
\end{figure}

The half-metallic behavior is once again shown in Fig. \ref{fig:Momentum_Spectral_function} where the momentum-dependent spectral functions close to the quantum phase transition are depicted.
While the spectrum for the spin-up channel (panel a) is almost indistinguishable from the spectrum in the SC phase at $J=0$ (not shown) and exhibits a gap,
the spectrum of the spin-down channel does not show any gaps and instead displays the properties of a metal.

Since the system can preserve the gap in one conduction band channel, the CDW+N\'eel state yields a small energy gain compared to the SC state
where the effective magnetic fields always decrease the size of the gap in both channels.
This opens the opportunity to use the combination of CDWs and N\'eel ordering,
which originates from the interplay between an attractive $U$ and a Kondo coupling $J$, 
as an application for spin filters.

For the momentum-dependent spectral functions of the magnetically ordered phase and the paramagnetic phase, 
we have not observed any differences compared to the standard Kondo lattice \cite{lit:Peters2015_2}.

\section{Away from Half Filling}
  \label{sec:away_from_half-filling}

\begin{figure}[t]
	\flushleft{(a)}
\includegraphics[width=0.49\textwidth]{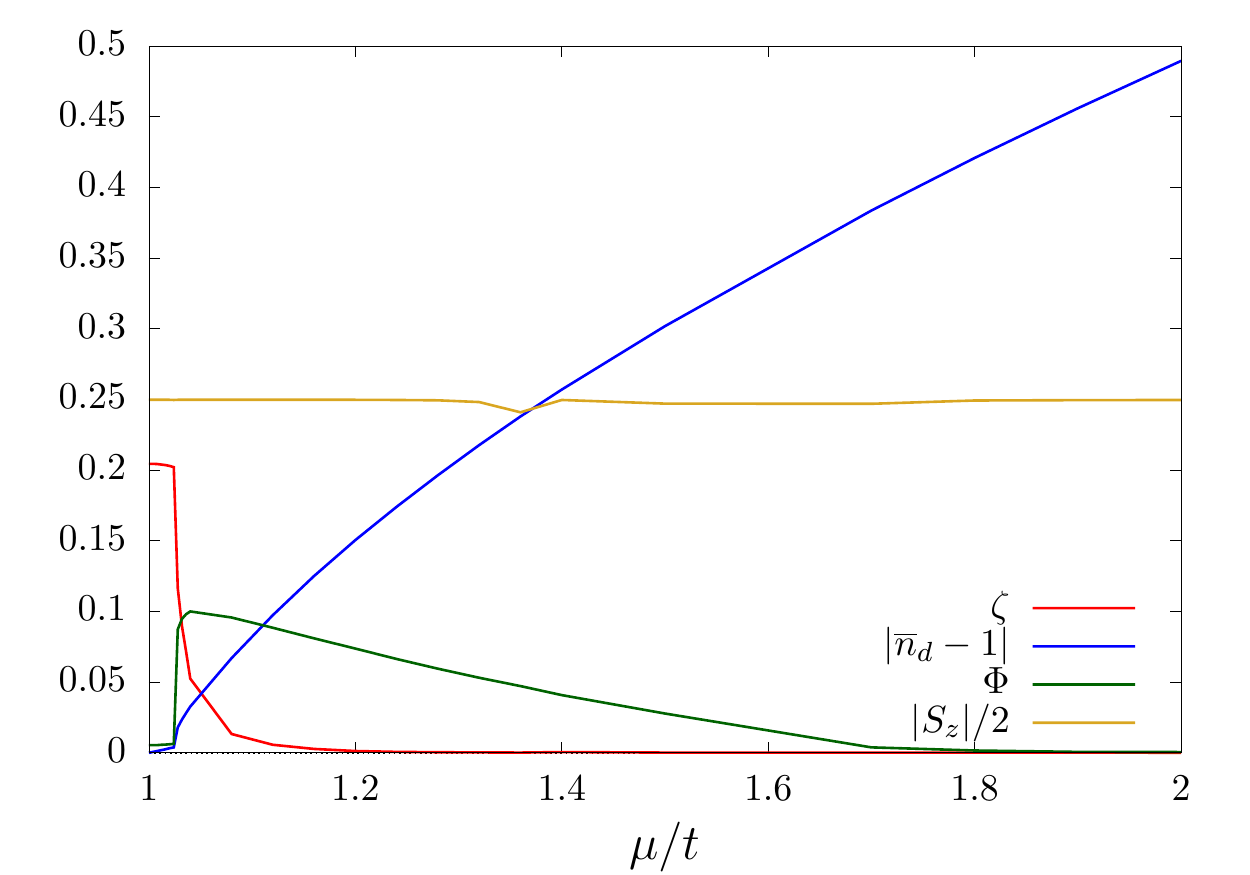}
	\flushleft{(b)}
	\includegraphics[width=0.49\textwidth]{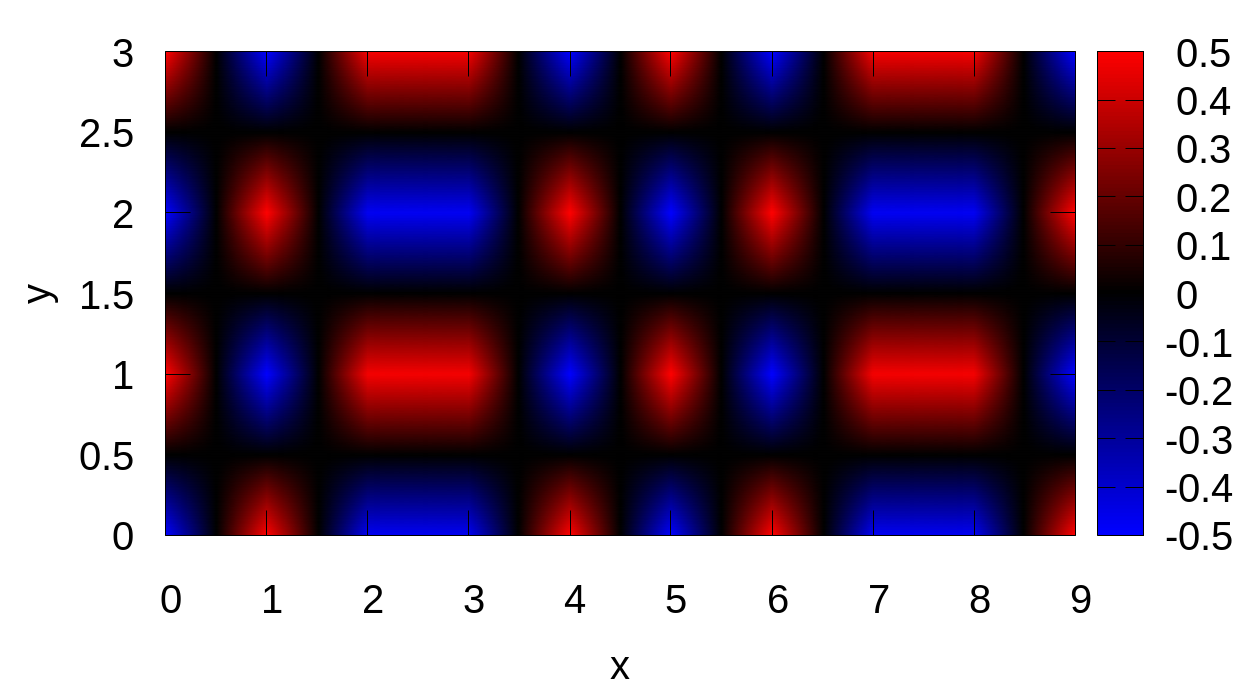}
	\caption{
		(a) CDW order parameter $\zeta$, averaged occupation $\overline{n}_d=1/N\sum_i n_{d,i}$, superconducting expectation value $\Phi$, and polarization of the localized spins $|S_z|$ 
		for $U/t=-2$ and $J/t=0.2$ as a function of the chemical potential $\mu$.
		For $\mu/t=1$, the lattice is half filled and CDWs occur.
		(b) Site-dependent polarization $S_z$ of the localized $f$-electron spins for $U/t=-2$, $J/t=0.2$ and $\mu/t=1.4$.
		Around $\mu/t=1.4$, the antiferromagnetic N\'eel state is not stable anymore and instead SDWs as shown in panel (b) appear.
		}
	\label{fig:EScan}
\end{figure}

In the attractive Hubbard model with $J=0$, the SC state and CDW state are degenerate only at half filling.
Away from half filling this degeneracy is lifted and instead only the SC state becomes the ground state.

Figure \ref{fig:EScan}(a) shows different properties of the system for $U/t=-2$ and a finite coupling $J/t=0.2$ as a function of the chemical potential $\mu$.

For the particle-hole symmetric case $\mu/t=1$, the average occupation number $\overline{n}_d=1$ indicates that the system is at half filling
and, consequently, we observe CDWs with $\zeta \neq 0$.

For a small critical deviation away from $\mu/t=1$ the CDW order parameter $\zeta$ shows a discontinuous jump to a value close to zero, indicating a first-order transition.
At the same time, also the SC expectation value jumps from $\Phi=0$ to a finite value.
At this point, the SC state, instead of the CDW + N\'eel state, becomes the new ground state.
The finite residual $\zeta$ is presumably caused by a finite energy resolution and broadening effects of the NRG spectra,
which have the same effect as a finite temperature in real experiments.

Using an applied voltage to change the chemical potential, it is, therefore, possible to drive the system from the insulating CDW phase at half filling to the SC phase.
This could be interesting for a possible future implementation of SC transistors.

Away from half filling, the superconductivity persists up to much larger couplings $J$ compared to the case of half filling.
The almost free $f$-electron spins are stabilized in a SDW state by the RKKY interaction.
In this phase, we observe superconductivity combined with magnetic ordering
confirming previous results \cite{lit:Bertussi2009,lit:Karmakar2016,lit:Costa2018}.
Upon further increasing the chemical potential, the small residual $\zeta$ rapidly disappears, 
leading to a complete breakdown of the CDWs while the deviation from half filling $|\overline{n}_d-1|$ continuously increases. 
The SC expectation value $\Phi$ decreases almost linearly with increasing $\mu$ until it vanishes around $\mu/t \approx 1.75$.

Away from half filling, however, the homogeneous N\'eel state becomes unstable and changes into a phase of SDWs as depicted in Fig. \ref{fig:EScan}(b)
in which the polarization of the localized spins are lattice-site dependent.
Exactly the same kind of SDWs have also been found for the normal Kondo lattice with $U=0$ \cite{lit:Peters2015_2,lit:Peters_2017}.
Although the superconductivity is still significant ($\Phi\approx 0.04$) in this regime, we can, therefore, conclude that it has no influence on the structure of the SDWs.
A finite attractive $U$ just changes the critical Kondo coupling at which the N\'eel state becomes unstable \cite{lit:Bertussi2009,lit:Costa2018}.

\begin{figure}[t]
	\flushleft{(a)}
	\includegraphics[width=0.49\textwidth]{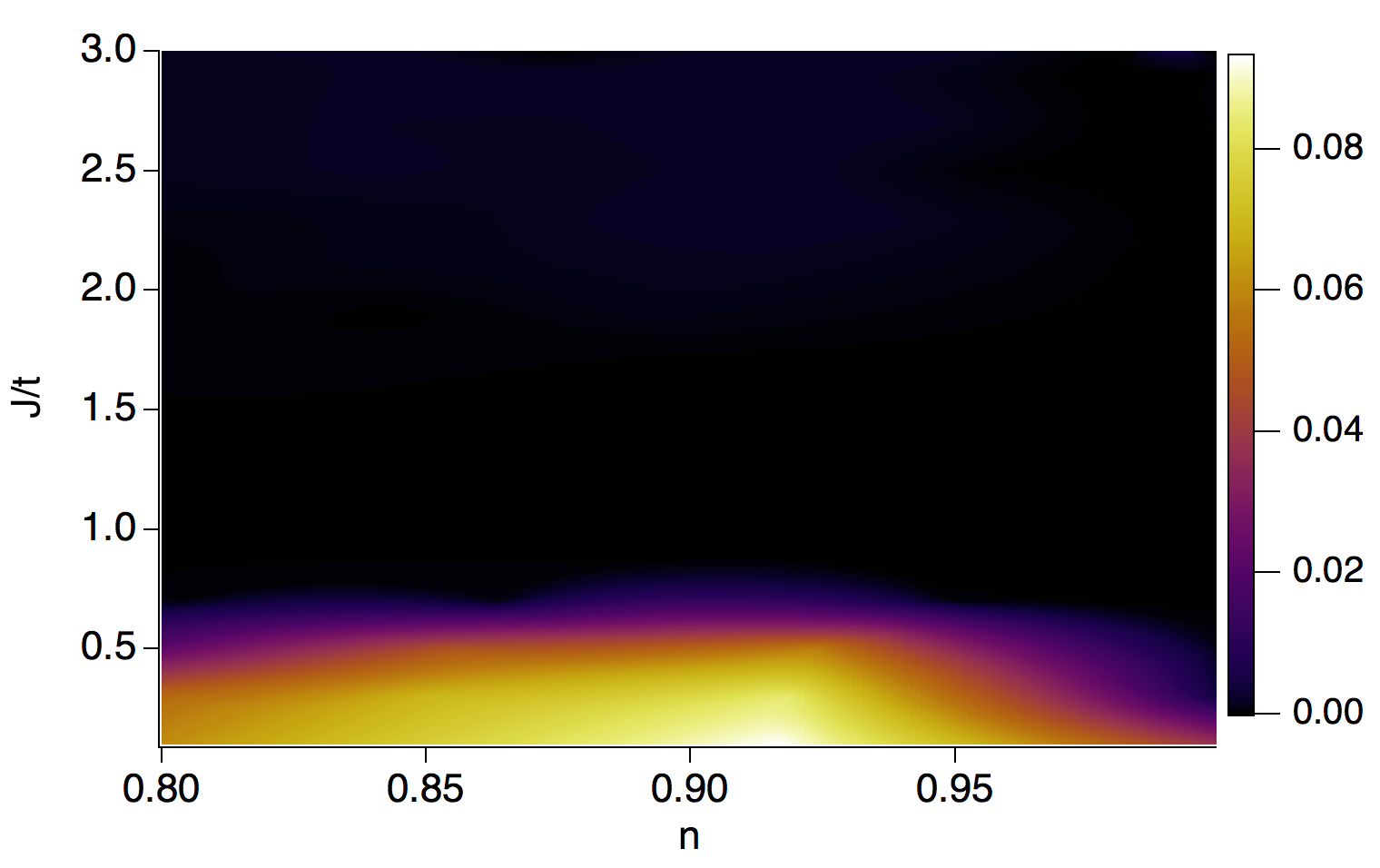}
	\flushleft{(b)}
	\includegraphics[width=0.49\textwidth]{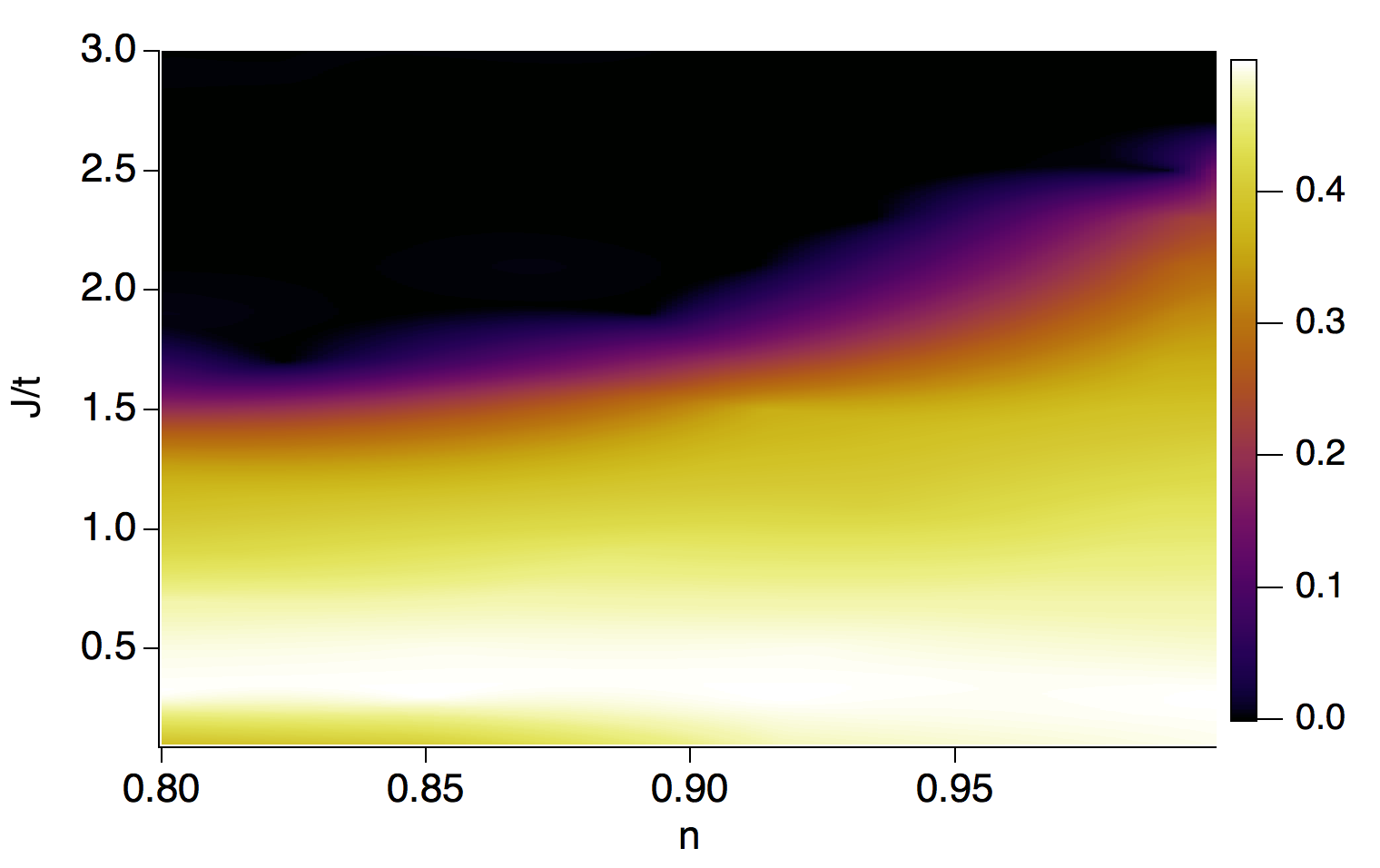}
	\caption{
		(a) Superconducting expectation value $\Phi$ and (b) polarization of the localized spins as a function of the coupling $J$ and the filling $n$
		for $U/t=-2$.
		}
	\label{fig:JScan_EScan}
\end{figure}

Figure \ref{fig:JScan_EScan} depicts the anomalous expectation value $\Phi$ and the polarization of the localized $f$-electron spins 
as a function of the coupling $J$ and the filling $n$ for $U/t=-2$.
Note that in contrast to the case of half filling, $\Phi$ continuously decreases with increasing Kondo coupling $J$ and no discontinuity occurs \cite{lit:Bertussi2009,lit:Karmakar2016,lit:Costa2018}.
Away from half filling, superconductivity can be observed for couplings up to $J/t \approx 0.5$
and it is largest for fillings around $n\approx 0.92$.
While for larger fillings than $n\approx 0.92$ superconductivity is suppressed since at half filling the CDW state is the ground state, 
for lower fillings it decreases because the electron density is reduced.

\begin{figure}[t]
	\flushleft{(a)}
	\includegraphics[width=0.49\textwidth]{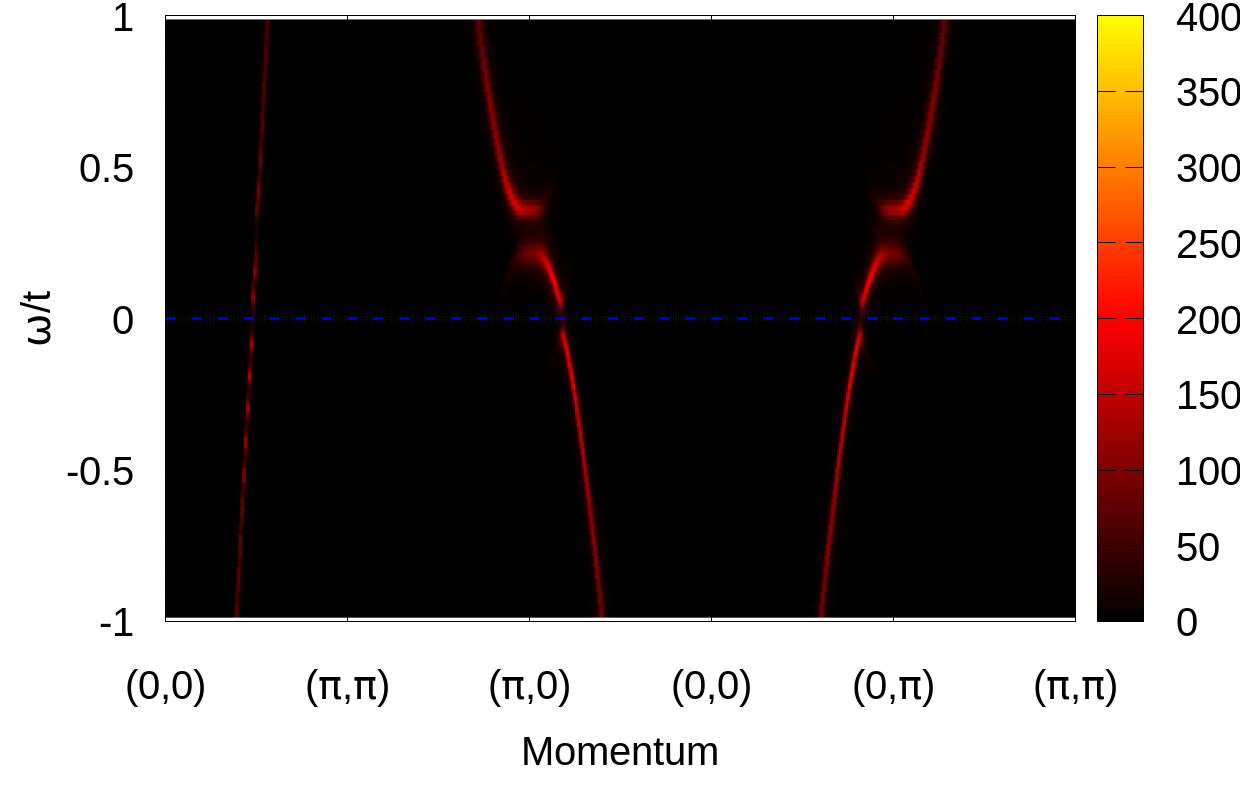}
	\flushleft{(b)}
	\includegraphics[width=0.49\textwidth]{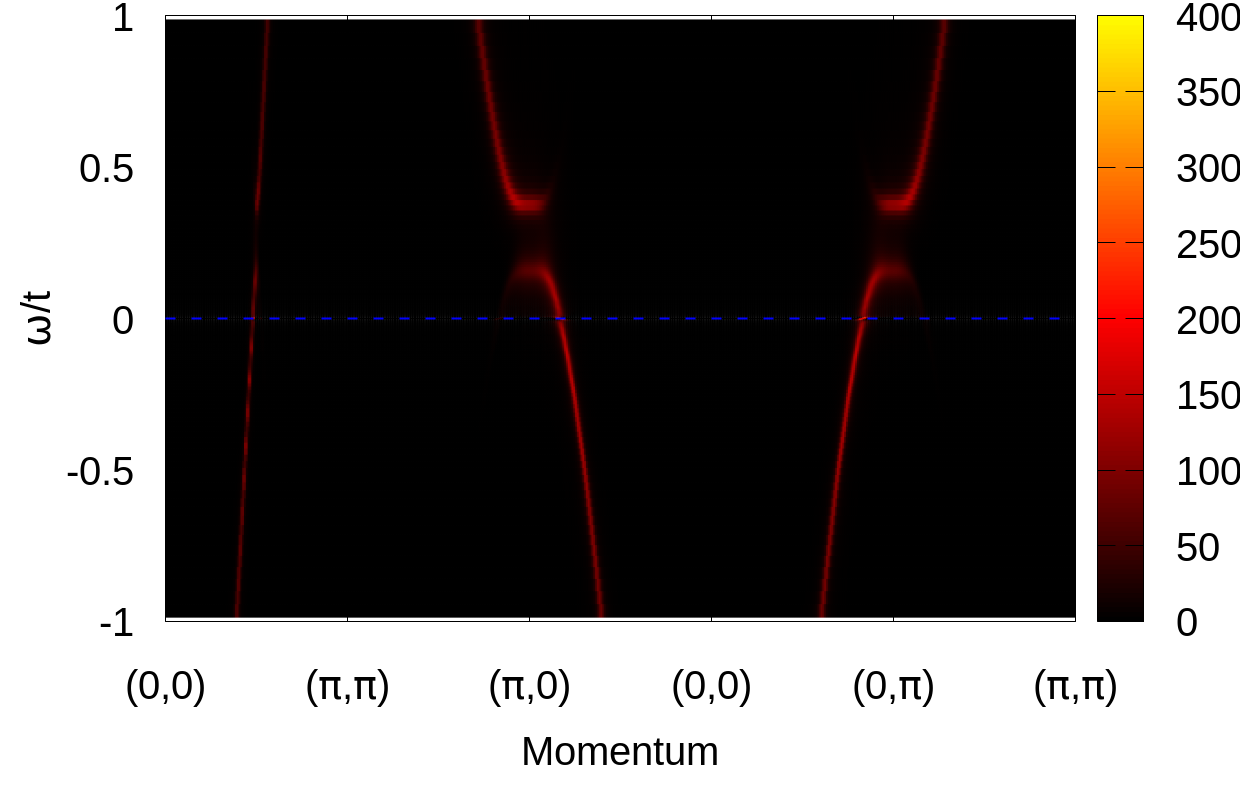}
	\caption{
		Moment-dependent spectral functions for (a) $J/t=0.5$ and (b) $J/t=0.7$ for $n \approx 0.85$ and $U/t=-2$.
		Blue dashed line indicates Fermi energy.
		}
	\label{fig:Doped_Spectrum_small_and_large_J}
\end{figure}

The momentum-dependent spectral functions for two different couplings $J/t=0.5$ and $0.7$ are shown in Fig. \ref{fig:Doped_Spectrum_small_and_large_J} 
for the filling $n\approx 0.85$ and $U/t=-2$.
For the smaller coupling $J/t=0.5$ [Fig. \ref{fig:Doped_Spectrum_small_and_large_J}(a)], the spectrum exhibits two gaps, one directly at the Fermi energy (indicated by a blue dashed line), and the other at $\omega/t=0.25$.
For this coupling strength, we still observe a significant anomalous expectation value of $\Phi \approx 0.05$ and
the gap at the Fermi energy is the SC gap.
This gap is largest for $J=0$ and becomes continuously smaller with increasing coupling $J$, which agrees with the observation that $\Phi$ continuously decreases with increasing $J$.

For the larger coupling $J/t=0.7$ [Fig. \ref{fig:Doped_Spectrum_small_and_large_J}(b)], the SC expectation value is zero $\Phi =0$ and, consequently, the gap at the Fermi energy is completely gone
so that the system behaves like a metal.
On the other hand, the width of the gap at $\omega/t=0.25$ is increased compared to the case for $J/t=0.5$.
This gap is already known from the ordinary Kondo lattice \cite{lit:Peters2015_2} and resides at half filling.
It is caused by the hybridization with the localized electrons and the width increases with increasing coupling $J$ \cite{lit:Coleman2006}.

\section{Conclusion}
  \label{sec:Conclusion}

In this paper, we have studied the competition between superconductivity, charge ordering, magnetic ordering, and the Kondo effect in a heavy fermion $s$-wave superconductor
which is described by the Kondo lattice model with an attractive on-site Hubbard interaction.
To solve this model, we have employed for the first time the combination of RDMFT and a newly developed self-consistent NRG scheme in Nambu space 
as an impurity solver.
Compared to the approach of Bauer \textit{et al.} \cite{lit:Bauer2009} we have chosen a different ansatz for the discretized impurity model
that allows SU(2) spin symmetry broken solutions, which is essential to study the competition between SDWs and superconductivity.

Using this new approach, we have found a rich phase-diagram at half filling, where depending on $J$ and $U$ many different effects may occur.
For very small Kondo couplings $J$ compared to the on-site interaction $U$, the system behaves like a Hubbard model with an attractive on-site interaction
while for large couplings the system shows the properties of a usual Kondo lattice with $U=0$.
For moderate couplings, we have found a completely new phase where CDWs and magnetic ordering are present at the same time.
Interestingly, the N\'eel state of the $f$-electron spins favors the CDW state over the SC state and, hence, lifts the degeneracy between the two phases such that superconductivity is strongly suppressed.
Another remarkable feature is that, in this phase, the system may become a half metal close to the quantum phase transition to the non-SC magnetically ordered phase
where the gap in the DOS closes only in one spin-channel of the conduction band.

Away from half filling, our findings are in good agreement with previous results \cite{lit:Bertussi2009,lit:Karmakar2016,lit:Costa2018}.
The CDWs are suppressed and we have found instead a phase where superconductivity along with magnetic ordering exists up to moderate couplings $J$.
For the chosen interaction $U/t=-2$, the superconductivity is strongest for fillings around $n\approx 0.9$.
The anomalous expectation value as well as the SC gap both decrease continuously with increasing coupling $J$.
Instead of the homogeneous N\'eel state, we have observed incommensurate SDWs.
Since the same kind of SDWs have already been seen in the ordinary Kondo lattice away from half filling \cite{lit:Peters2015_2}, 
we find no evidence that superconductivity has an influence on the structure of these SDWs.
A finite attractive $U$ just changes the Kondo coupling, at which the incommensurate SDWs occur.

Since an applied voltage can change the chemical potential and drive the system from the insulating CDW state at half filling to the SC state away from half filling,
this system might be interesting for a possible future implementation of a SC transistor, where superconductivity can be switched on and off simply by applying a voltage.

Interestingly, superconductivity away from half filling as well as CDW at half filling both enhance the magnetic ordering since the gap in the DOS mitigates the Kondo screening \cite{lit:Raczkowski2010}
such that the $f$-electron spins are almost completely polarized in both phases.

In future work, our enhanced RDMFT+NRG approach could be used to investigate a variety of other SC systems
since it is not limited to homogeneous SC lattice systems, where localized $f$-electrons reside on every lattice site.
One example could be diluted SC systems where the behavior for different impurity concentrations is examined.
In this case, one would randomly place a specific number of impurities on the lattice sites of a large RDMFT cluster such that the desired concentration is achieved.
On the other hand, one could also study proximity-induced superconductivity where a lattice Hubbard model with an attractive on-site potential $U$ 
is coupled, e.g., to an ordinary Kondo lattice.
These issues are now under consideration.

\begin{acknowledgments}
	B.L. thanks the Japan Society for the Promotion of Science (JSPS) and the Alexander von Humboldt Foundation.
	Computations were performed at the Supercomputer Center, Institute for Solid State Physics, University of Tokyo and the Yukawa Institute for Theoretical Physics, Kyoto.
	This   work   is   partly   supported   by   JSPS   KAKENHI Grants No.  JP15H05855, JP16K05501, JP17F17703, JP18H01140,
	JP18K03511, and No.  JP18H04316.
\end{acknowledgments}

%

\end{document}